\documentclass{article}

\usepackage{arxiv}

\usepackage[utf8]{inputenc} 
\usepackage[T1]{fontenc}    
\usepackage{hyperref}       
\usepackage{url}            
\usepackage{booktabs}       
\usepackage{amsfonts}       
\usepackage{nicefrac}       
\usepackage{microtype}      
\usepackage{graphicx}
\usepackage{natbib}
\usepackage{doi}
\usepackage{colortbl}  
\usepackage[table,xcdraw]{xcolor}  
\usepackage{graphicx}   
\usepackage{subcaption} 
\usepackage{amsmath}
\usepackage{multirow} 
\usepackage{threeparttable}  

\title{A Robust Multi-Stage Intrusion Detection System for In-Vehicle Network Security using Hierarchical Federated Learning}

\author{ \href{https://orcid.org/0000-0002-2150-7981}{\includegraphics[scale=0.06]{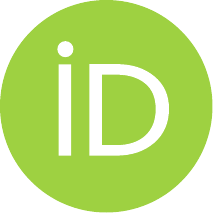}\hspace{1mm}Muzun Althunayyan} \\
	School of Computer Science \& Informatics\\
	Cardiff University\\
	Cardiff, United Kingdom \\
	\texttt{AlthunayyanMS@cardiff.ac.uk} \\
	\And
	\href{https://orcid.org/0000-0001-9761-0945}{\includegraphics[scale=0.06]{orcid.pdf}\hspace{1mm}Amir Javed} \\
	School of Computer Science \& Informatics\\
	Cardiff University\\
	Cardiff, United Kingdom \\
	\texttt{javeda7@cardiff.ac.uk} \\
 \And
	\href{https://orcid.org/0000-0003-3597-2646}{\includegraphics[scale=0.06]{orcid.pdf}\hspace{1mm}Omer Rana} \\
	School of Computer Science \& Informatics\\
	Cardiff University\\
	Cardiff, United Kingdom \\
	\texttt{ranaof@cardiff.ac.uk} 
}

\begin{document}
\maketitle

\begin{abstract}
As connected and autonomous vehicles proliferate, the Controller Area Network (CAN) bus has become the predominant communication standard for in-vehicle networks due to its speed and efficiency. However, the CAN bus lacks basic security measures such as authentication and encryption, making it highly vulnerable to cyberattacks. To ensure in-vehicle security, intrusion detection systems (IDSs) must detect seen attacks and provide a robust defense against new, unseen attacks while remaining lightweight for practical deployment. Previous work has relied solely on the CAN ID feature or has used traditional machine learning (ML) approaches with manual feature extraction. These approaches overlook other exploitable features, making it challenging to adapt to new unseen attack variants and compromising security. This paper introduces a cutting-edge, novel, lightweight, in-vehicle, IDS-leveraging, deep learning (DL) algorithm to address these limitations. The proposed IDS employs a multi-stage approach: an artificial neural network (ANN) in the first stage to detect seen attacks, and a Long Short-Term Memory (LSTM) autoencoder in the second stage to detect new, unseen attacks. To understand and analyze diverse driving behaviors, update the model with the latest attack patterns, and preserve data privacy, we propose a theoretical framework to deploy our IDS in a hierarchical federated learning (H-FL) environment. Experimental results demonstrate that our IDS achieves an F1-score exceeding 0.99 for seen attacks and exceeding 0.95 for novel attacks, with a detection rate of 99.99\%. Additionally, the false alarm rate (FAR) is exceptionally low at 0.016\%, minimizing false alarms. Despite using DL algorithms known for their effectiveness in identifying sophisticated and zero-day attacks, the IDS remains lightweight, ensuring its feasibility for real-world deployment. This makes our model robust against seen and unseen attacks.
\end{abstract}


\section{Introduction}
\label{Introduction}
In connected and autonomous vehicles (CAVs), many mechanical components have been replaced by electronic components \cite{foster2015exploring}. These vehicles contain numerous Electronic Control Units (ECUs) connected via various standard automobile in-vehicle communication protocols, such as Controller Area Network (CAN), FlexRay, Local Interconnect Network (LIN), and Media Oriented System Transport (MOST). Among these protocols, CAN is considered the de facto protocol for in-vehicle communication \cite{al2019intrusion} due to its features including high speed, noise cancellation, and ease of use. Although it was initially developed for industrial machines, it has since been adopted for in-vehicle network communications. However, it lacks basic security features such as sender authentication and encryption \cite{paul2021artificial}. The main reasons for not implementing these security measures in in-vehicle networks are the intensive use of the vehicle’s limited computational resources and the resultant increased latency, which could potentially lead to a failure to meet critical safety-related deadlines \cite{pese2021s2, barati2014energy}. Thus, any security measure designed should be lightweight to ensure ease of deployment.

Furthermore, interconnectivity in modern vehicles introduces attack surfaces that expose the vehicle to cyberattacks. The attack surfaces can be accessed physically or remotely \cite{checkoway2011comprehensive}. Physical access can be made via USB, CD player, onboard diagnostic (OBD)-II port, and so on. In addition, remote access can be made via short-range wireless technologies such as Bluetooth and radio frequency identification (RFID) and long-range wireless technologies such as Wi-Fi and long-term evolution (LTE). Therefore, the system is vulnerable to various cyberattacks, potentially resulting in severe consequences, including the loss of human life \cite{young2019survey}. There may be adverse consequences if an intruder manages to infiltrate the CAN bus system and introduce malicious messages. For instance, an unauthorized individual with access to the in-vehicle network can tamper with vital functionalities like braking, door locking mechanisms, and steering, thereby presenting a notable safety hazard. An example of a cybersecurity breach in the automotive industry can be found in \cite{golson2016jeep}, where two hackers were able to exploit vulnerabilities in a Jeep’s system and remotely control it to perform dangerous maneuvers, including abruptly turning the steering wheel and suddenly applying the parking brake at high speeds, leading to catastrophic accidents. In the same way, hackers were able to exploit a vulnerability in the infotainment system of a General Motors car, allowing them to gain unauthorized access and steal data remotely \cite{crume2015ownstar}. In 2018, the Keen Security Lab revealed several vulnerabilities in BMW cars that allow cyber-attackers to inject unified diagnostic services (UDS) packets into the CAN network, evading the central gateway \cite{tencentexperimentalBMW}. Moreover, in 2020, a Toyota Lexus was the target of an attack by exploiting a Bluetooth vulnerability, causing unexpected physical motions in the vehicle \cite{tencentexperimental}. 

Such attacks not only raise concerns about information security and privacy but also directly impact the safety of drivers, passengers, and the surrounding area. Therefore, the security of the CAN bus has become a topic of significant research interest. According to McKinsey’s analysis in \cite{bertoncello2021unlocking}, almost all newly sold vehicles worldwide, approximately 95\%, will feature connectivity by 2030, underscoring the critical need for robust CAN bus security measures. It has been proven that IDSs are an effective method for identifying cyberattacks on in-vehicle networks. An IDS monitors and identifies malicious activity on the network. In the context of in-vehicle networks, the IDS is often installed in an ECU and receives and analyzes incoming network traffic. If any abnormal messages are identified, it will notify other ECUs. In computer network systems, IDSs are used to detect and prevent intrusions. However, many conventional network security measures cannot be applied directly to in-vehicle networks. Therefore, there is a critical need for a robust IDS for in-vehicle networks.

Many studies have developed various in-vehicle IDSs using machine learning (ML) approaches. However, existing approaches overlook three crucial aspects of in-vehicle IDS requirements: robustness, limited computing resources, and deployment environment. Robustness in an IDS is the ability to detect seen attacks while staying ahead of attackers by identifying new, unseen attacks that the model has not been trained on. Ensuring the robustness of the IDS will significantly enhance the safety and security of in-vehicle networks. To design lightweight IDSs deployable in resource-constrained environments, many existing solutions rely solely on CAN ID features or use traditional ML approaches with manual feature extraction, neglecting other exploitable features and thus compromising security. Moreover, various researchers have deployed their IDSs using a traditional centralized learning approach, which involves transmitting large volumes of data to the cloud for training. This approach raises significant issues, including privacy concerns, high communication overhead, and longer response times. This paper aims to address these gaps by proposing a robust and lightweight in-vehicle IDS capable of defending against both seen and new, unseen attacks, and which can be deployed in a federated learning (FL) environment. To do so, we have laid down concrete objectives:
\begin{itemize}
    \item Identify the limitations of existing in-vehicle IDSs and design a novel IDS to address these gaps.
    \item Understand CAN bus data to develop a hybrid IDS.
    \item Conduct data preprocessing for DL  applications.
    \item Evaluate the hybrid IDS for both seen and unseen attacks, considering the model size.
    \item Compare the model's performance against previous studies.
    \item Propose a hierarchical federated learning (H-FL) IDS to increase the robustness of the IDS.
\end{itemize}

\subsection{Contributions}
This paper proposes a novel multi-stage IDS for in-vehicle network security capable of detecting both seen and unseen attacks. Furthermore, to take advantage of diverse driving scenarios and behaviors across different locations while ensuring user privacy protection, we propose a novel theoretical framework for deploying the proposed IDS in an H-FL environment. The main contributions can be summarized as follows: 
\begin{itemize}
    \item To provide a literature review on the current research on ML approaches capable of detecting seen and unseen attacks, and to identify the limitations of existing in-vehicle IDSs.
    
    \item It examines CAN bus data to extract valuable insights, aiding researchers in enhancing their understanding and improving security solutions in this field. 

    \item It proposes a robust multi-stage IDS to detect seen and unseen malicious traffic using DL algorithms, achieving an F1-score exceeding 0.95 and a detection rate (DR) of 99.99\% for previously unseen attacks. 
    
    \item We have developed a novel method for detecting unseen attacks by utilizing an LSTM-autoencoder. This approach captures the sequential dependencies within a single CAN message, encompassing both the CAN ID and its payload. 
    
    \item Focusing on the size, the proposed novel hybrid model is lightweight at 2.98 MB, demonstrating its feasibility for real-world deployment.
    
    \item This paper proposes a novel framework for deploying the proposed IDS in an H-FL environment.  
\end{itemize}

This paper advances the field of in-vehicle IDSs by addressing key limitations of existing solutions. It leverages the power of DL algorithms and integrates two stages of detection to enhance robustness, deploying the IDS in an H-FL environment. These improvements set a new standard for further advancements toward achieving an optimally secure in-vehicle network, making it more difficult for attacks to succeed. Moreover, our analysis of CAN bus data provides valuable insights for researchers, aiding in the development of more effective security measures for in-vehicle networks. Our work paves the way for future research to explore advanced DL models, ultimately enhancing the security and reliability of CAVs. By pushing existing boundaries, we seek to contribute to a securer and more resilient automotive cybersecurity landscape. To the best of our knowledge, this is the first study to utilize a hybrid approach in in-vehicle IDSs to detect seen and unseen attacks using DL algorithms rather than traditional ML algorithms and to propose deploying the in-vehicle IDS in an H-FL.

The remainder of this paper is organized as follows. Section \ref{Background} presents the background of the CAN bus protocol and attack methods. Section \ref{Related_Work_Section} discusses the related work. Section \ref{Methodology_Section} presents the methodology of the proposed IDS. Section \ref{Results_Section} presents and evaluates the experimental results. Section \ref{Discussion} provides discussions. Finally, Section \ref{Conclusion} concludes the paper and presents potential directions for future work.

\section{Background and Attack Methods}
\label{Background}
In this section, we provide background information on CAN bus protocol security and attack methods.

\subsection{Controller Area Network Protocol}
\label{CANBackground}
The CAN bus is the primary communication protocol between multiple ECUs in in-vehicle networks. Robert Bosch developed the CAN bus protocol in 1985 to reduce the weight of wires, complexity, and cost \cite{specification1991bosch}. Due to its high speed and efficiency, the CAN bus is widely used as the default communication standard for in-vehicle communication systems \cite{hoang2022detecting}. The CAN bus is a message-based broadcast protocol where ECUs transmit data in a predefined data frame as messages to all ECUs. Despite its importance, the CAN bus protocol lacks security features in its design, making it vulnerable to attacks on confidentiality, integrity, and availability \cite{paul2021artificial, rajapaksha2023ai, aliwa2021cyberattacks}.

\begin{figure}[h!t]
\centering
\begin{subfigure}{0.28\textwidth}
    \includegraphics[width=\textwidth]{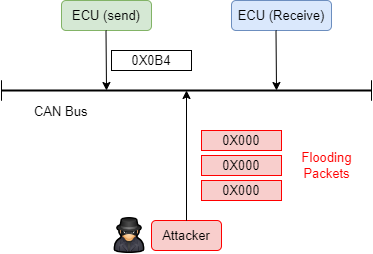}
    \caption{DoS Attack}
    \label{fig:flooding}
\end{subfigure}
\hfill
\begin{subfigure}{0.28\textwidth}
    \includegraphics[width=\textwidth]{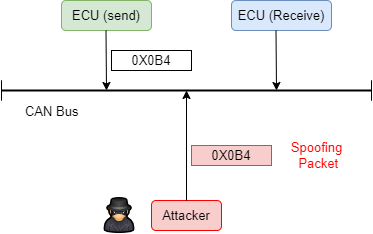}
    \caption{Spoofing Attack}
    \label{fig:spoofing}
\end{subfigure}
\hfill
\begin{subfigure}{0.28\textwidth}
    \includegraphics[width=\textwidth]{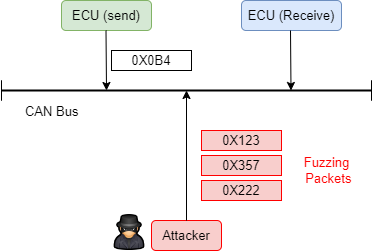}
    \caption{Frame Fuzzification Attack}
    \label{fig:fuzzy}
\end{subfigure}
\hfill
\caption{CAN Bus Attacks}
\label{fig:CAN Bus Attacks}
\end{figure}

\subsection{Attack Methods}
As mentioned in the \ref{CANBackground} subsection, the CAN bus is vulnerable to various cyberattacks due to its broadcast nature and lack of built-in encryption or authentication \cite{paul2021artificial}. To launch an attack, the attacker must first gain access to the CAN bus system, which is possible through different methods like the OBD-II port or wireless technologies \cite{aliwa2021cyberattacks,checkoway2011comprehensive}. Having successfully gained access, the attacker can inject harmful messages and launch various attacks such as DoS, spoofing, and frame fuzzification attacks. These attacks are depicted in Figure \ref{fig:CAN Bus Attacks}.

\textbf{Denial-of-Service Attack (DoS):} The goal of a DoS attack is to consume the CAN bus bandwidth by transmitting a large volume of messages, which may result in unexpected system behavior. As the identifier sets the message priority, the attacker sends many messages with identifier = 0 to the CAN bus (see Figure ~\ref{fig:flooding}). This message will have the highest priority \cite{hoang2022detecting}.

\textbf{Spoofing Attack:} In this type of attack, an unauthorized attacker targets specific CAN IDs and tries to inject fabricated messages to control particular functions (see Figure ~\ref{fig:spoofing}). Because the CAN IDs are spoofed and look legitimate, it becomes challenging to distinguish between legitimate and fabricated messages, leading to system malfunction \cite{hoang2022detecting}.

\textbf{Frame Fuzzification Attack:} An attacker injects random messages, which appear to be legitimate traffic, into the CAN bus network system (see Figure ~\ref{fig:fuzzy}). Frame fuzzification attacks can compromise the ECUs and cause unexpected behavior in the vehicle, leading to problems like steering wheel shaking, unpredictable signal light on/off switching, and automatic gear shifts \cite{lee2017otids}.  

\section{Related Work}
\label{Related_Work_Section}
This section reviews state-of-the-art research focused on the development of in-vehicle IDSs capable of detecting both seen and unseen attacks, and their limitations.  It also reviews existing federated-based in-vehicle IDSs, examines the broader impact of in-vehicle network security on urban transportation systems, and highlights the novelty of our research within this context.

\subsection{In-Vehicle IDSs and Limitations}
Research on developing in-vehicle IDSs has significantly increased over the past few years due to the critical need to enhance the security of in-vehicle networks and identify cyberattacks. Researchers have explored a wide range of approaches to developing these systems. In this paper, we focus on studies that develop in-vehicle IDSs using DL and ML approaches. An IDS can be either a signature-based or an anomaly-based system \cite{hoppe2009applying}. Some studies have focused on signature-based methods to detect and classify seen attacks \cite{paul2021artificial, hossain2020lstm, hossain2020effective}, while others have employed anomaly detection methods to identify unseen or zero-day attacks \cite{song2021self, wei2022novel}. To further improve the robustness and detection capability of the in-vehicle IDS, a few papers have developed an in-vehicle IDS that can identify both seen and unseen attacks \cite{zhang2019intrusion, hoang2022detecting, seo2018gids, MTH-IDS2022}, demonstrating significant advancements in this critical area of cybersecurity. These studies and their limitations are discussed below and summarized in Table \ref{Related Work}.

\begin{table*}[h!t]
\centering
\caption{Related Works of ML-based IDSs for in-vehicle Network}
\label{Related Work}
\resizebox{\textwidth}{!}{
\begin{tabular}{cclcccccccccc}
\hline
\textbf{Ref} & \textbf{Year} & \textbf{Category} & \textbf{Algorithm} & \textbf{Dataset} & \textbf{\begin{tabular}[c]{@{}l@{}}Seen \\ Attacks\end{tabular}} & \multicolumn{1}{c}{\textbf{\begin{tabular}[c]{@{}c@{}}Unseen\\ Attacks\end{tabular}}} & \textbf{M-C}& \multicolumn{1}{c}{\textbf{\begin{tabular}[c]{@{}c@{}}ID-based\\ Detection\end{tabular}}} &
\multicolumn{1}{c}{\textbf{\begin{tabular}[c]{@{}c@{}}Payload-based\\ Detection\end{tabular}}} &\textbf{DL} &\textbf{\begin{tabular}[c]{@{}l@{}}FL \end{tabular}} \\ \hline
\\[-.8em]
\cite{zhang2019intrusion} & 2019 & Supervised & DNN & Simulation & \checkmark & \checkmark  &  & \checkmark  & \checkmark & \checkmark& \\ \hline
\\[-.8em]
\cite{hoang2022detecting} & 2022  & \begin{tabular}[c]{@{}c@{}}Semi-\\supervised \end{tabular} & \begin{tabular}[c]{@{}c@{}}AE, \\ GAN \end{tabular} &Car-Hacking \cite{seo2018gids} & \checkmark & \checkmark  &  & \checkmark  & & \checkmark \\ \hline
\\[-.8em]
\cite{seo2018gids} & 2018  & Unsupervised &  GAN & Car-Hacking \cite{seo2018gids} & \checkmark & \checkmark  &  & \checkmark  & &\checkmark & \\ \hline
\\[-.8em]

\cite{MTH-IDS2022}  & 2022 & Hybrid & \begin{tabular}[c]{@{}c@{}}DT, RF, ET, \\ XGBoost, \\CL-k-means \end{tabular}  & \begin{tabular}[c]{@{}c@{}}Car-Hacking \cite{seo2018gids},\\ CICIDS2017  \cite{sharafaldin2018toward}\end{tabular}  & \checkmark &  \checkmark & \checkmark & \checkmark & \checkmark &  & \\ \hline
\\[-.8em]
\rowcolor{yellow!50}
\begin{tabular}[c]{@{}c@{}} \textbf{Our} \\\ \textbf{work} \end{tabular} &  \textbf{2023} &  \textbf{Hybrid} &\begin{tabular}[c]{@{}c@{}} \textbf{ANN}, \\\textbf{LSTM-AE} \end{tabular}  &  \textbf{Car-Hacking} \cite{seo2018gids} & \checkmark &  \checkmark &  \checkmark & \checkmark & \checkmark & \checkmark & \checkmark\\ \hline
\\[-.8em]
\end{tabular}}
  \begin{tablenotes}
       \small
       \item \textbf{DL:} Deap Learning, \textbf{FL:} Federated Learning, \textbf{M-C:} Multi-classification. 
    \end{tablenotes}
\end{table*}

Zhang et al. \cite{zhang2019intrusion} have proposed a DNN-based IDS that aims to automatically extract features for the IDS from the vehicle’s data packets. The authors applied gradient descent with momentum (GDM) and gradient descent with momentum and adaptive gain (GDM/AG) techniques. The study’s results demonstrate the model’s capability to detect replay attacks effectively. The authors accessed the sensor readings, using them as separate features. However, this method demands either having access to the CAN database (DBC) file or having knowledge about the CAN payload. 

Hoang et al. \cite{hoang2022detecting} and Seo et al. \cite{seo2018gids} have showcased their IDSs’ ability to detect both seen and unseen attacks. However, their IDSs mainly rely on the CAN ID as a singular feature; selecting only the CAN ID feature will limit the detection ability to detect attacks that involve payload manipulation \cite{rajapaksha2023ai}.

Hoang et al. \cite{hoang2022detecting} propose a lightweight, semi-supervised, learning-based IDS to detect in-vehicle network attacks. The proposed IDS in their study integrates two DL models: autoencoders and generative adversarial networks (GAN). Their IDS was trained on unlabeled data to learn the patterns of normal and malicious data. Only a few labeled samples were used during the subsequent supervised training phase.  

Seo et al. \cite{seo2018gids} have developed a GAN-based IDS (GIDS) for in-vehicle network security. The proposed IDS was trained by solely utilizing the patterns of CAN IDs from CAN data and then converting the extracted CAN IDs into a simple image. GIDS has two discriminative models to detect both seen and unseen attack data. The first discriminator is specifically trained to handle attacks. In contrast, the second discriminator and the generator are co-trained through an adversarial process. While the generator generates modified images, the second discriminator receives both modified and real CAN images, and its role is to differentiate between the modified and real images. Nevertheless, using the CAN ID as the only feature limits the detection capability of payload manipulation attacks. 

The aforementioned studies employ either supervised learning-based methods or unsupervised learning-based methods. Supervised learning-based methods are frequently used to develop models distinguishing between normal and attack traffic. This technique focuses on optimizing the decision boundary to minimize classification errors on the training set and demonstrate good generalization capabilities on the testing set. Despite the high accuracy and low false alarm rate (FAR) of these models, they rely heavily on well-labeled and balanced datasets to reach their full potential, which is often difficult to achieve in practical situations. Additionally, because attackers continuously attempt to evade detection and use new, previously unseen attacks, supervised learning-based models may struggle to recognize unfamiliar attack patterns not present in the training data \cite{vikram2020anomaly}. This limitation can lead to significant security consequences. Unsupervised learning-based methods, meanwhile, build models exclusively using normal data, relying on profiling normal traffic behaviors to detect anomalous traffic that could indicate a potential attack. As a result, unsupervised learning-based models are well suited to detecting previously unseen attacks \cite{pratomo2018unsupervised}. 

To leverage the strengths of both approaches, Yang et al. \cite{MTH-IDS2022} have developed a multi-tiered hybrid IDS, MTH -IDS, to protect intra-vehicle and external networks from cyberattacks. MTH-IDS combines supervised and unsupervised models. Despite achieving good results, the proposed IDS has certain limitations that we aim to address. 

One significant limitation to note is that the authors used only four features—CAN ID, D5, D3, and D1—to train the model after feature extraction. While this strategy may result in a more efficient model, it does create the potential for an attacker to manipulate other features that were not considered during the model's training process \cite{kocher2021machine}. This is a crucial limitation in CAN bus data for three reasons. First, selecting a subset of features from the CAN bus payload as important features and discarding non-important features could pose a risk, as attackers might exploit the neglected features, effectively bypassing the model \cite{li2014feature, zhang2015adversarial}. Second, the evolving landscape of attack scenarios means that features chosen to detect one category of attacks may become outdated or insufficient for addressing new, unseen attacks \cite{kocher2021machine}. Third, upon closer examination of the CAN bus data, we noticed that each CAN ID exhibits distinct patterns of data field values. For example, CAN ID 399 consistently has zero values in data fields D2, D3, D4, D6, and D7, which remain unchanged. By contrast, in CAN ID 320, the data fields that are always 0 are D1, D2, and D3. Table \ref{Constant_Values} displays the constant zero values in blue, indicating data that never changes throughout the dataset. Data highlighted in yellow represents changing values, while data highlighted in green denotes constant non-zero values. Consequently, features that are significant (possessing varying values) for one CAN ID may not hold relevance for another, making the selection of a consistent subset of important features for all CAN IDs impractical due to the inherent nature of CAN bus data. This observation can help researchers understand the CAN bus data and make their IDS models more focused on the most significant features, potentially improving their generalization capability.  

Another limitation pertains to the deployment approach. Yang et al. suggest conducting the training process on a server machine and the testing process on the CAN bus. However, this deployment approach requires training the data on a remote server, which involves sending sensitive data to a server for training, raising privacy concerns. 

In the unsupervised model, the F1-score of the proposed MTH-IDS was only 0.82. To improve this result, the authors add another tier of two biased classifiers. Training these biased classifiers on specific errors—false positives (FPs) and false negatives (FNs)—may lead to poorer performance when we test the model on new, unseen data. Additionally, adding this tier transforms the model from being purely unsupervised. Resulting in a dependency on labeled datasets, which are often difficult to implement in practical situations.

A key distinction between MTH-IDS and our approach is our use of DL algorithms over conventional ML models. Multiple authors have found that DL-based IDSs outperform traditional ML IDSs in automotive applications \cite{mehedi2021deep}. This superiority is due to several factors: DL methods are more adaptive, continually being refined with incoming data, which is particularly suitable for the nature of CAN bus data \cite{zhang2019intrusion}. Additionally, traditional ML often requires manual feature engineering, such as applying correlation-based feature selection, which can be time-consuming \cite{nagarajan2023machine}. In contrast, DL automatically deduces features, allowing algorithms to directly discern optimal features from raw data \cite{lampe2023survey}. Furthermore, DL IDSs are especially capable of detecting novel attacks and can scale more effectively to highly complex in-vehicle network data while maintaining efficacy \cite{lampe2023survey}. 
Therefore, there is a need to develop a robust, lightweight in-vehicle IDS that addresses previous limitations.

\begin{table}[]
\centering
\caption{Constant and Changing Data in CAN Message}
\label{Constant_Values}
\begin{tabular}{c*{8}{p{.45cm}}}
\hline
\textbf{CAN ID} & \textbf{D0} & \textbf{D1} & \textbf{D2} & \textbf{D3} & \textbf{D4} & \textbf{D5} & \textbf{D6} & \textbf{D7} \\ \hline
399 & \cellcolor[HTML]{A6E2BE}254 & \cellcolor[HTML]{FFFFC7}59 & \cellcolor[HTML]{DAE8FC}00 & \cellcolor[HTML]{DAE8FC}00 & \cellcolor[HTML]{DAE8FC}00 & \cellcolor[HTML]{FFFFC7}60 & \cellcolor[HTML]{DAE8FC}00 & \cellcolor[HTML]{DAE8FC}00 \\ \hline
399 & \cellcolor[HTML]{A6E2BE}254 & \cellcolor[HTML]{FFFFC7}60 & \cellcolor[HTML]{DAE8FC}00 & \cellcolor[HTML]{DAE8FC}00 & \cellcolor[HTML]{DAE8FC}00 & \cellcolor[HTML]{FFFFC7}61 & \cellcolor[HTML]{DAE8FC}00 & \cellcolor[HTML]{DAE8FC}00 \\ \hline
\textbf{\begin{tabular}[c]{@{}c@{}}. \\ . \\ .\end{tabular}} & \cellcolor[HTML]{A6E2BE}\textbf{\begin{tabular}[c]{@{}c@{}}. \\ . \\ .\end{tabular}} & \cellcolor[HTML]{FFFFC7}\textbf{\begin{tabular}[c]{@{}c@{}}. \\ . \\ .\end{tabular}} & \multicolumn{1}{l}{\cellcolor[HTML]{DAE8FC}\begin{tabular}[c]{@{}l@{}}00\\ 00\\ 00\end{tabular}} & \multicolumn{1}{l}{\cellcolor[HTML]{DAE8FC}\begin{tabular}[c]{@{}l@{}}00\\ 00\\ 00\end{tabular}} & \multicolumn{1}{l}{\cellcolor[HTML]{DAE8FC}\begin{tabular}[c]{@{}l@{}}00\\ 00\\ 00\end{tabular}} & \cellcolor[HTML]{FFFFC7}\textbf{\begin{tabular}[c]{@{}c@{}}. \\ . \\ .\end{tabular}} & \multicolumn{1}{l}{\cellcolor[HTML]{DAE8FC}\begin{tabular}[c]{@{}l@{}}00\\ 00\\ 00\end{tabular}} & \multicolumn{1}{l}{\cellcolor[HTML]{DAE8FC}\begin{tabular}[c]{@{}l@{}}00\\ 00\\ 00\end{tabular}} \\ \hline
399 & \cellcolor[HTML]{A6E2BE}254 & \cellcolor[HTML]{FFFFC7}94 & \cellcolor[HTML]{DAE8FC}00 & \cellcolor[HTML]{DAE8FC}00 & \cellcolor[HTML]{DAE8FC}00 & \cellcolor[HTML]{FFFFC7}74 & \cellcolor[HTML]{DAE8FC}00 & \cellcolor[HTML]{DAE8FC}00 \\ \hline
399 & \cellcolor[HTML]{A6E2BE}254 & \cellcolor[HTML]{FFFFC7}95 & \cellcolor[HTML]{DAE8FC}00 & \cellcolor[HTML]{DAE8FC}00 & \cellcolor[HTML]{DAE8FC}00 & \cellcolor[HTML]{FFFFC7}75 & \cellcolor[HTML]{DAE8FC}00 & \cellcolor[HTML]{DAE8FC}00 \\ \hline
\end{tabular}
\end{table}

\subsection{Federated-based In-Vehicle IDSs}
FL is a privacy-preserving decentralized learning technique that trains models locally without transferring row data to a centralized server. Instead, it transfers model parameters to a centralized server, which aggregates the clients' models to build a shared global model \cite{agrawal2022federated}. The integration of FL into IDSs addresses the growing need for enhanced security and privacy in our interconnected society. Although some previous works \cite{alsamiri2023federated,driss2022federated,shibly2022personalized, yu2022federated, zhang2023federated, yang2022federated} have deployed their in-vehicle IDSs in an FL environment, all the proposed FL works used a standard (non-hierarchical) architecture, relying solely on one central server (aggregator). However, this can introduce performance challenges, particularly due to delays in sharing the model between the central aggregator and numerous devices. Furthermore, this central server poses a risk as a single point of failure \cite{rana2023hierarchical} and is restricted to limited driving behaviors under its coverage  \cite{liu2020client}.

\subsection{Impact on Urban Transportation Systems}
The impact of securing in-vehicle networks in CAVs extends to the broader field of urban transportation planning, where vehicles are integral to the overall system. Urban transportation research aims to mitigate traffic congestion \cite{almatar2023traffic}, improve safety, increase the adoption of renewable energy \cite{almatar2023increasing}, and advance sustainable mobility \cite{almatar2023towards}. A recent survey \cite{rahman2023impacts} examining the impacts of CAVs on urban transportation and the environment found that CAVs would decrease energy consumption and protect the environment by reducing emissions. Furthermore, these vehicles could significantly reduce traffic crashes involving human error while increasing the convenience and productivity of passengers. However, there are widespread concerns about personal safety, security, and privacy due to the potential for cyberattacks. Therefore, enhancing the security of CAVs would significantly contribute to the improvement of urban transportation systems. 

\subsection{Research Novelty}
While previous studies have achieved notable results in specific areas, they also have several limitations. Compared to existing studies related to in-vehicle IDS, our proposed IDS offers the following advantages: 1) It uses DL rather than traditional ML, which has proven to be more efficient in automotive applications and is capable of detecting novel attacks more effectively \cite{mehedi2021deep, lampe2023survey}. 2) It employs a hybrid model (signature-based and anomaly detection) instead of relying on a single model, which has been shown to improve detection results \cite{musa2020intrusion}. 3) It successfully detects both seen and new, unseen attacks. 4) Despite using DL algorithms, it is a lightweight IDS. 5) It utilizes both CAN ID and payload as features, which enables detection of CAN ID changes and payload manipulation attacks \cite{rajapaksha2023ai}. 6) It continuously learns by labeling and updating the signature-based classifier when a new, unseen attack is detected. 7) It leverages diverse driving behaviors by being deployed in an H-FL environment. 

The novelty of this paper lies in three key aspects: the design of the IDS, the algorithms used, and the deployment approach. First, our proposed in-vehicle IDS employs a cascaded multi-stage approach to detect both seen and unseen attacks. Unlike other multi-stage IDS designs in the literature, our IDS uses the first stage to classify traffic into seen attacks and normal data. The second stage then re-examines the data classified as normal in the first stage to detect unseen attacks that bypassed the initial model, providing an additional layer of protection. Furthermore, the IDS continuously learns by labeling and updating the classifier in the first stage when a new, unseen attack is detected. Secondly, we use a hybrid approach (ANN and LSTM-autoencoder) that leverages DL algorithms. By doing so, our proposed IDS ensures a multi-layered defense mechanism against potential threats and improves detection performance compared to single-point IDSs \cite{hbaieb2022federated}. Thirdly, to further enhance the robustness of the in-vehicle IDS, our approach addresses the significant drawbacks of previous methods by proposing its deployment in an H-FL environment. This approach aims to build a more robust global model that takes advantage of diverse driving scenarios and behaviors in different locations while ensuring user privacy protection.

In summary, the review of existing literature reveals several research gaps and shortcomings, highlighting the broader impact of in-vehicle network security on urban transportation systems and emphasizing the need to improve the security of these networks. These gaps provide a critical foundation for the design and novelty of our proposed IDS.

\section{Methodology}
\label{Methodology_Section}
In this section, we present the methodologies used to develop our proposed in-vehicle IDS. First, we explain the proposed in-vehicle IDS, including data preprocessing; the first stage, a supervised model (ANNs); and the second stage, an unsupervised model (LSTM-autoencoder). Next, we provide details on training and testing the model, including ANN hyperparameter tuning and LSTM-autoencoder hyperparameter tuning. Finally, we present the proposed H-FL framework.

\subsection{Proposed In-vehicle IDS}
In Section \ref{Related_Work_Section}, we highlighted the limitations of existing in-vehicle IDSs. To address these limitations, we introduce a multi-stage IDS designed to protect in-vehicle networks from seen and unseen attacks by using a hybrid approach (supervised and unsupervised algorithms). We adopt a hybrid approach to mitigate the risks inherent in relying on a single model.  Integrating multiple ML models can significantly enhance performance, improve data security, and reduce the FN rate compared to using a single model \cite{driss2022federated, musa2020intrusion}. These benefits are particularly valuable for in-vehicle networks, where errors can be costly. Additionally, our proposed hybrid approach increases protection against attackers, who must now evade two models instead of just one. 

\begin{figure*}[h!t]
\centering
    \includegraphics[scale=.15]{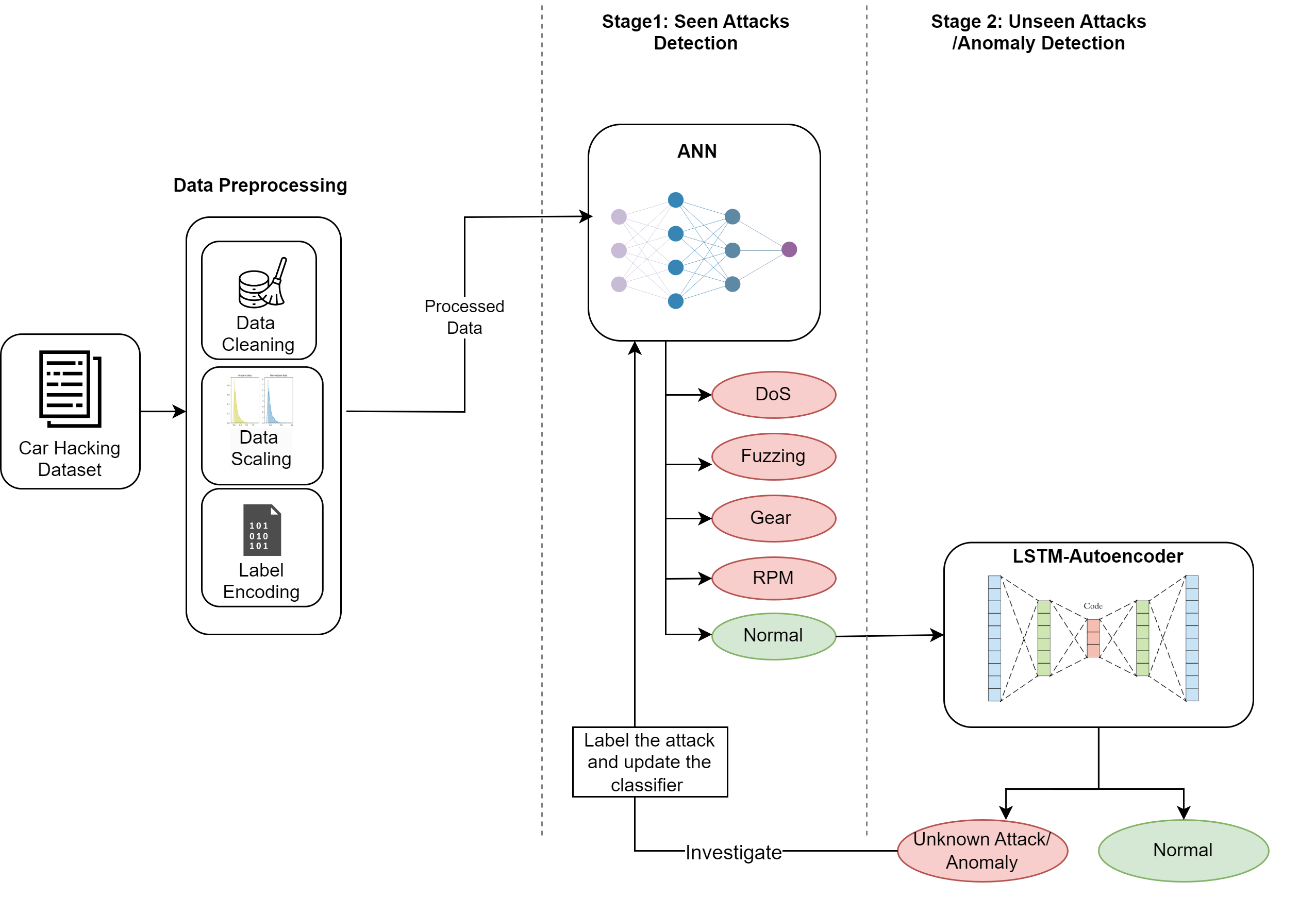}
    \caption{Workflow Model Depiction of Proposed Multi-stage IDS for In-vehicle Network}
    \label{fig:Workflow}
\end{figure*}

Our proposed IDS integrates both supervised and unsupervised models. As illustrated in Figure \ref{fig:Workflow}, the CAN bus data first enters the data preprocessing stage, followed by the supervised classifier in the initial stage. Subsequently, the normal data is processed by the second model to identify any previously unseen attacks. The supervised model is primarily responsible for detecting and categorizing previously seen attacks based on historical data. By placing the supervised classifier model first, the IDS can quickly filter out any attacks based on its training and accelerate the detection process. The subsequent unsupervised model serves as a secondary layer of protection against unseen attacks that bypass the first model. When the supervised model makes a mistake and classifies the malicious traffic as normal, the unsupervised model can detect this and flag it as an anomaly. The unsupervised model, which works as an anomaly detection model, is trained solely on normal data, and any samples that deviate significantly from the learned patterns are identified as an anomaly or an unseen attack. Once the unsupervised model detects malicious traffic, it is flagged for further investigation. Any anomalies detected by the unsupervised model that are later confirmed as threats will have a new attack label generated. This new label will be used to further train and refine the supervised model, enabling it to recognize such attacks in the future and ensuring the system improves over time. As the vehicular environment evolves, new attack vectors may emerge. The unsupervised model ensures that the system remains adaptive and resilient in the face of changing attacks, even if the supervised model has not been trained on them. This comprehensive multi-stage IDS ensures coverage for both seen and unseen attacks. For broader applicability, the proposed IDS can learn the legitimate CAN IDs and normal behavior for each vehicle at design time and monitor the network to detect any attacks during operational runtime.

Our proposed IDS provides a comprehensive and robust solution to secure in-vehicle networks. Furthermore, our IDS addresses significant challenges associated with previous approaches by deploying it in a FL environment. Utilizing FL allows the IDS to benefit from various driving scenarios, enhances its resilience against new and unseen attacks, and enables continuous learning without compromising the privacy and security of the training data. With the rapid development of in-vehicle technologies and communication protocols, having a resilient IDS ensures that we are prepared for both current and future threats.

\subsubsection{Data Preprocessing}
\label{Data_Preprocessing}

\begin{figure}[h!t]
\centering
    \includegraphics[scale=.20]{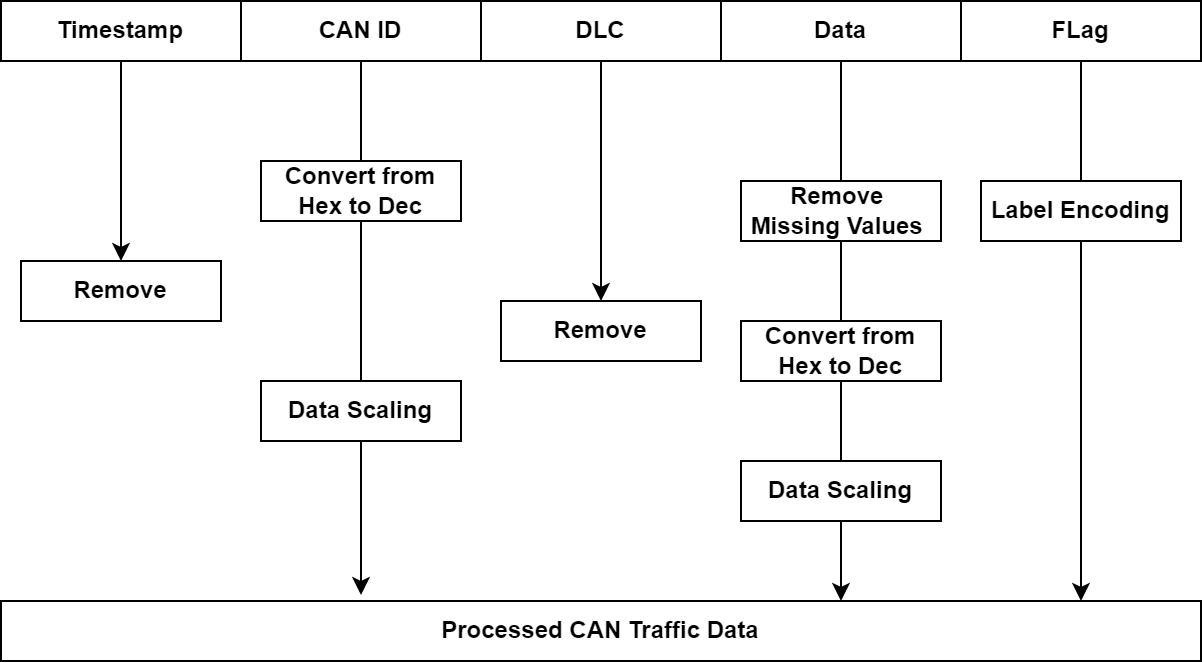}
    \caption{Data Preprocessing}
    \label{fig:DataPreprocessing}
\end{figure}

In data preprocessing, data was converted into a format suitable for use by deep neural networks. This was achieved by applying different operations. Figure \ref{fig:DataPreprocessing} shows the data processing applied to each feature in the dataset. The dataset contained four files, each corresponding to a specific attack (DoS, frame fuzzification, gear, and RPM), with both attack and normal instances. Table \ref{DatasetOverview} shows the number of attacks and normal instances in each file. We processed each file individually and then concatenated them into a single dataset. We shifted the flag field to the last column and filled non-available data bytes with NAN  values. Given the extensive data points we had, we removed any row with these missing values in the data fields. The CAN ID and data fields were in hexadecimal values. We converted the CAN ID and data values from hexadecimal to decimal values as per the ML requirement. Most ML algorithms require the conversion of categorical data into numerical data. Therefore, after concatenating the datasets into one dataset, a label encoder was employed to convert categorical features in the flag feature (Normal, DoS, frame fuzzification, gear, and RPM) into numerical representations to facilitate their use as inputs in ML algorithms.

\begin{figure}[h!t]
\centering
    \includegraphics[scale=.55]{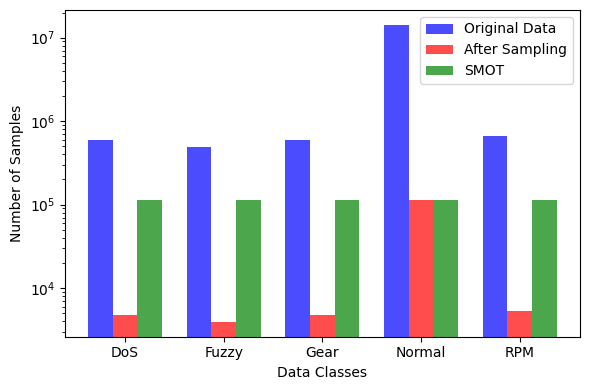}
    \caption{Comparison between the number of data classes before and after sampling and balancing data}
    \label{Samples}
\end{figure}

As shown in Table \ref{DatasetOverview}, the dataset comprised millions of data points and was highly imbalanced, with the number of attacks accounting for 14.06\%  and normal data making up 85.94\%. This could result in high processing times and produce biased models. To mitigate issues related to the dataset’s size and model complexity, data sampling is a common approach used to generate a representative sample from the original data \cite{faraoun2006neural}. After reviewing and experimenting with various data sampling methods, we adopted the same sampling approach used in \cite{MTH-IDS2022}. In their work, they employed the K-means clustering algorithm for data sampling and then addressed class imbalance issues using the synthetic minority oversampling technique (SMOTE). Figure \ref{Samples} depicts the difference between the number of samples in the original data and the number after applying sampling and SMOTE to balance the data. Although we trained the models on sampled data, we tested them on the remaining dataset to ensure that the models were well generalized and capable of recognizing the entire dataset. We used this method only for unsupervised anomaly detection because the supervised classifier handled the entire dataset efficiently and quickly.

\begin{table}[ht!]
\centering
\caption{Dataset Overview}
\label{DatasetOverview}
\begin{tabular}{cccc}
\hline
\multicolumn{1}{l}{\textbf{Attack type}} & \textbf{Attack Instances} & \textbf{Normal Instances}\\ \hline \\[-.8em]
DoS & 587,521  & 3,078,250 \\ \hline \\[-.8em]
Frame Fuzzification & 491,847  & 3,347,013 \\ \hline \\[-.8em]
Gear & 597,252  & 3,845,890  \\ \hline \\[-.8em]
RPM & 654,897  & 3,966,805 \\ \hline \\[-.8em]
Total & 2,331,517 (14.06\%) & 14,237,958 (85.94\%)\\ \hline \\[-.8em]
\end{tabular}
\end{table}

We considered nine columns (CAN ID and eight data values) as the features in our experiment. This choice was made to effectively detect any manipulation of either the CAN ID or the payload. We opted to include the CAN ID feature because the utilized dataset \cite{seo2018gids} was created by injecting attacks into arbitrary CAN IDs, which made incorporating this feature crucial for capturing the distinct characteristics of cyberattacks. To avoid biased models and potentially inaccurate results due to the clear correlation between timestamps and cyberattack simulation intervals, the timestamp feature was intentionally excluded. In addition, we excluded the data length code (DLC) feature since it could be correlated with the data fields, given that DLC represents the payload’s length. To decrease the feature dimensions, we decided not to consider DLC. We then applied the StandardScaler algorithm to scale the data. In comparison to other algorithms, StandardScaler was able to find a better balance between handling outliers and preserving the range of values. It standardizes data by subtracting the mean value, resulting in a zero mean, and then dividing by the variance, thus giving the distribution unit variance, as shown in Equation (\ref{eq:scaler}): 

\begin{equation} 
\label{eq:scaler}
\ X_{\text{scaled}} = ({X-X_{mean}})/{X_{std}}
\end{equation}

Where, $X_{\text{scaled}}$ represents the scaled value, $X$ is the original value, and $X_{\text{mean}}$ and $X_{\text{std}}$ denote the mean and standard deviation, respectively.

This step is important because network traffic data can possess differing ranges, and normalized (scaled) datasets tend to enhance the performance of ML models \cite{ali2018intelligent}. Failure to normalize the dataset, especially when its features have different scales, may cause the ML model to concentrate primarily on the features with larger scales\cite{MTH-IDS2022}. Table \ref{DatasetBefore_AfterPreprocess} shows the data features and types before and after data preprocessing.

\begin{table}[ht!]
\centering
\caption{Data Features and Types Before and After Data Preprocessing}
\label{DatasetBefore_AfterPreprocess} 
\begin{tabular}{lll}
\hline \\[-.8em]
\textbf{Feature} & \textbf{Before} & \textbf{After} \\ \hline \\[-.8em]
CAN ID & hexadecimal & decimal integer \\ \hline \\[-.8em]
D0 & hexadecimal & decimal integer \\ \hline \\[-.8em]
D1 & hexadecimal & decimal integer \\ \hline \\[-.8em]
D2 & hexadecimal & decimal integer \\ \hline \\[-.8em]
D3 & hexadecimal & decimal integer \\ \hline \\[-.8em]
D4 & hexadecimal & decimal integer \\ \hline \\[-.8em]
D5 & hexadecimal & decimal integer \\ \hline \\[-.8em]
D6 & hexadecimal & decimal integer \\ \hline \\[-.8em]
D7 & hexadecimal & decimal integer \\ \hline \\[-.8em]
Flag & string (T, R) & decimal integer (0,1,2,3,4)\\ \hline \\[-.8em]
\end{tabular}
\end{table}

\subsubsection{First Stage: Supervised Model}
We reviewed various DL algorithms for attack detection and multi-classification and concluded that artificial neural networks (ANNs) were the optimal choice for our data for the following reasons:

\begin{enumerate}
    \item ANNs introduce non-linearity into the model, which makes them capable of modeling complex patterns and relationships that linear classifiers might miss.
    \item Unlike traditional methods, ANNs can automatically learn the best features directly from the raw data without requiring explicit feature engineering. This can lead to better performance, especially in CAN bus data, where each CAN ID has distinctively informative features.
    \item ANNs are adaptive systems, making them suitable in situations where data is continuously evolving.
    \item ANNs can produce highly competitive results.
\end{enumerate}

However, while ANNs can be an excellent choice, they can be computationally intensive. To address this, we simplified the ANN architecture by reducing the number of layers and neurons while still achieving the highest DR.

\subsubsection{Artificial Neural Networks}
\begin{figure}[h!t]
\centering
    \includegraphics[scale=.15]{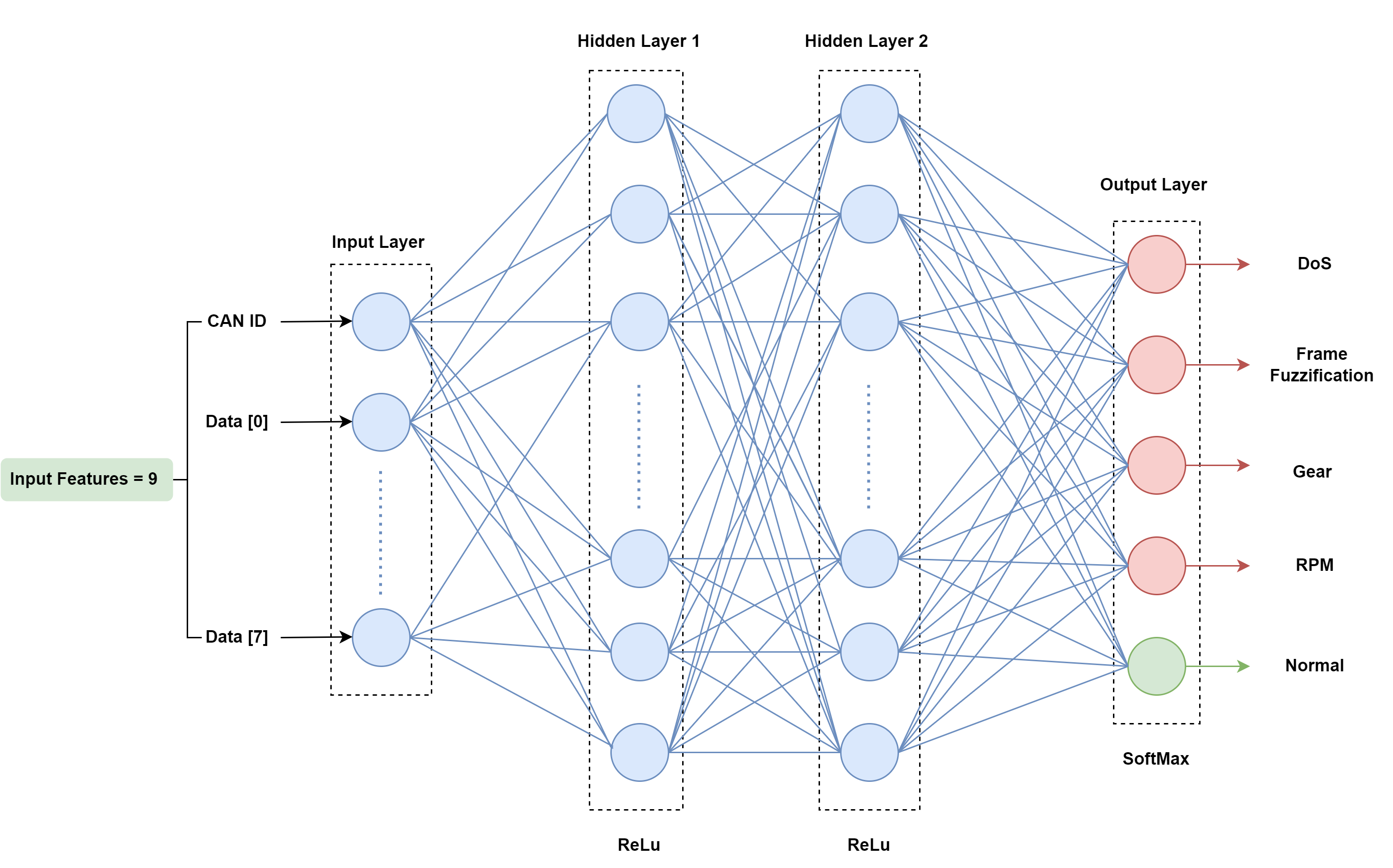}
    \caption{ANN Architecture}
    \label{fig:ANN architecture}
\end{figure}

ANNs are ML algorithms inspired by the behavior of biological neurons in the brain and the central nervous system \cite{mcculloch1943logical, rosenblatt1958perceptron}. ANNs’ inputs pass through one or more hidden layers, assign weights, and produce an output. Figure \ref {fig:ANN architecture} depicts the ANN architecture. ANNs can adjust their internal parameters, known as weights and biases, for both the hidden and output layers. This adaptive feature means that ANNs can understand the deep and non-linear interrelations between dependent and independent variables without any prior knowledge \cite{tu1996advantages}. In contrast to traditional classification algorithms that often demand knowledge of the system’s probability model, ANNs operate as a “black box” that can adapt to the underlying system model \cite{zhang2000neural}. Their adaptability, particularly in high-dimensional datasets, addresses challenges associated with conventional algorithms such as k-nearest neighbor and decision trees \cite{dreiseitl2002logistic}. Across various application domains, ANNs have been employed for classification tasks and have also demonstrated their efficacy in several computer security areas, including detecting network attacks \cite{wu2015network, shenfield2018intelligent}.

\subsubsection{Second Stage: Unsupervised Model}
The second stage of the proposed multi-stage IDS is an unsupervised model that acts as an anomaly detector, identifying anomalies or unseen attacks that bypass the first stage. Unsupervised learning-based models are particularly well suited to detecting previously unseen attacks \cite{pratomo2018unsupervised}. We selected the LSTM-autoencoder model as the anomaly detector model for in-vehicle IDS for several reasons. Firstly, as an unsupervised neural network, the autoencoder does not require labeled data, saving significant time and effort. Secondly, research shows that autoencoders are effective at detecting anomalies and unseen attacks, making them well suited for in-vehicle IDS \cite{hoang2022detecting, wei2022novel, INDRA2020}. Lastly, LSTM layers capture temporal dependencies within sequential data, a feature that conventional autoencoders lack. These features make the LSTM-autoencoder model a robust and efficient solution for detecting anomalies and enhancing the security of in-vehicle networks.

\subsubsection{LSTM-Autoencoder}
The Long Short-Term Memory (LSTM) autoencoder consists of both LSTM and autoencoder components. LSTM, a type of recurrent neural network (RNN), is designed to handle sequential data and can learn complex dynamics within the temporal order of input sequences by using internal memory to store information over long sequences. This capability is particularly useful for CAN bus data, which is sequential \cite{rajapaksha2023ai}.
In contrast, autoencoders are neural network architectures designed to learn efficient representations of input data by attempting to reconstruct the original data as accurately as possible. By combining LSTM with an autoencoder, we aimed to capture the sequence order in each CAN bus message, which classic autoencoders might overlook. The chosen LSTM-autoencoder consists of two interconnected LSTMs: the first encodes sequences of features into a fixed-size vector, and the second decodes this vector back into a sequence. 
In the context of detecting anomalies in the CAN bus, the LSTM-autoencoder was trained only on normal data. This allowed it to accurately learn to reconstruct the benign patterns it was trained on. When test data was presented, input data was first encoded by the LSTM into a fixed-size vector. Another LSTM then decoded this vector to reconstruct the input data. Any deviations in reconstruction can be potential indicators of anomalies.

\begin{figure}[h!t]
\centering
    \includegraphics[scale=.18]{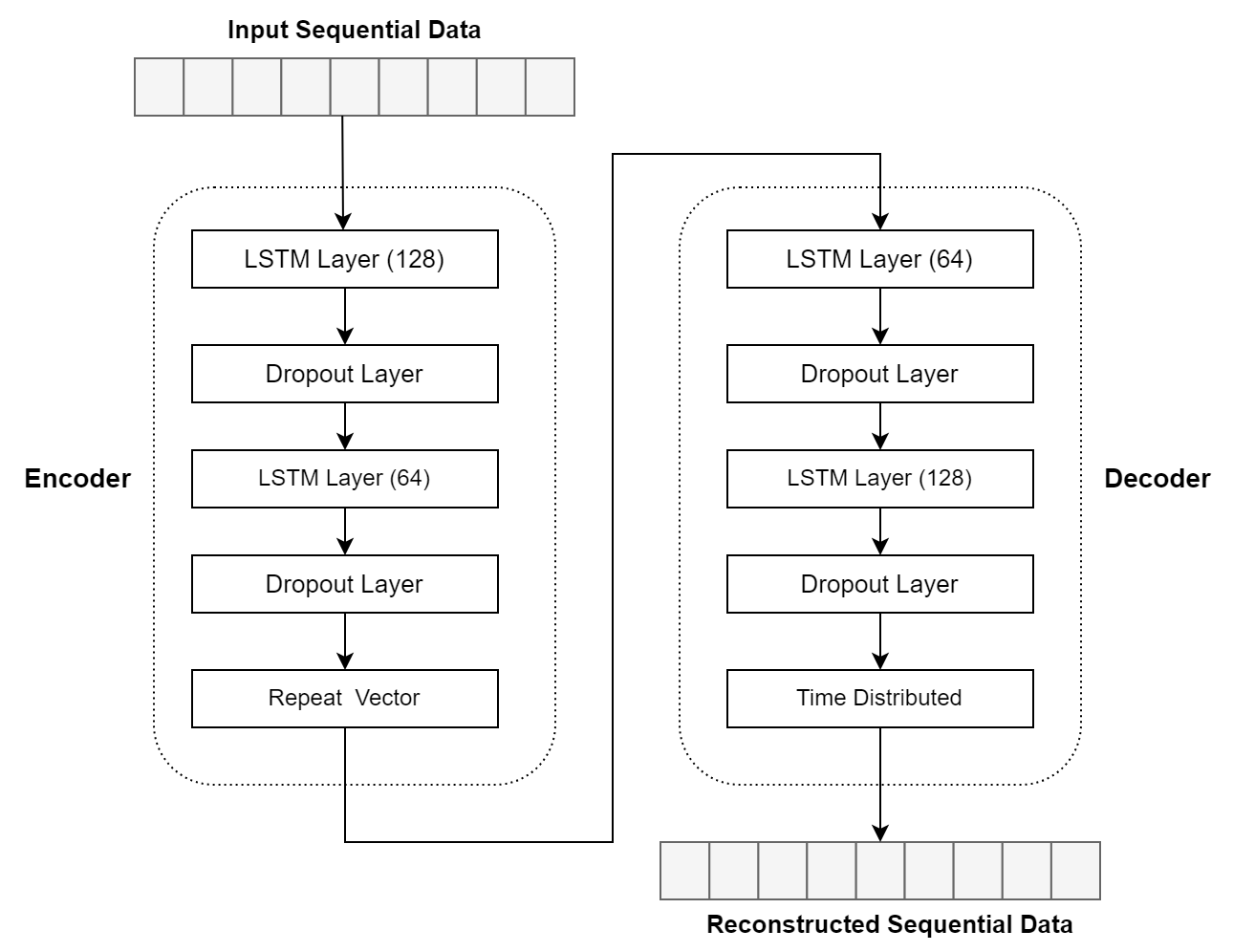}
    \caption{LSTM-Autoencoder Architecture}
    \label{fig:LSTM Architecture}
\end{figure}

The motivation for using the LSTM-autoencoder in this context arose from the sequential nature of the CAN ID and its associated data fields. Incorporating both the CAN ID and its data fields(payload) is essential because normal data fields for one CAN ID might be considered unusual for another CAN ID. The LSTM-autoencoder was trained on normal sequences, with any deviation from these established sequences subsequently flagged as anomalous. However, it is worth noting that several studies have used sequence-based models to explore the sequential dependencies between CAN IDs only or multiple CAN bus messages to detect anomalies. However, to the best of our knowledge, there has been no research examining the order dependence within a single CAN message, including both the CAN ID and its payload.  This is important because CAN IDs can follow periodic or event-driven patterns, and sequences of CAN IDs can change when event-triggered messages occur on the CAN bus \cite{rajapaksha2023ai}. For example, the sequence of CAN IDs or CAN messages can change when a sudden event happens, such as someone opening the door.
We trained an LSTM-autoencoder model to reconstruct benign sequences with minimal errors, expecting the model’s inputs and outputs to look alike. However, when a malicious sequence was fed into the model, the model was expected to fail at reconstructing the sequence. Therefore, the input and output vectors were expected to differ significantly. Figure \ref{fig:LSTM Architecture} depicts the architecture of the LSTM-autoencoder.

\subsection{Training and Testing the Model}
To prevent overfitting in ML models, we applied the data partitioning method outlined in \cite{gao2019intrusion} by dedicating 70\% to training and the subsequent 30\% to testing the model’s performance. In the ANN model, we trained the model using labeled data to learn the relationships between the features and the targets. Then, we tested the model on test data to assess the quality of the model. The ANN works as a multi-class classifier, and the output will be normal or attack type, including DoS, frame fuzzification, RPM, and gear spoofing. Further details regarding these attacks are discussed in Section \ref{DatasetDescription}. The data classified as normal in the first stage of the IDS (ANN) served as input for the second stage (the LSTM-autoencoder).
Our experiment involved adjusting multiple hyper-parameters, such as batch size, number of units, learning rates, optimizers, activation, and loss functions. Through systematic experimentation, we identified the most suitable hyper-parameter values that could deliver the best possible performance and efficiency.

\subsection{ANN Hyperparameter Tuning}
The model was a feedforward ANN that comprised four layers: an input layer, two hidden layers, and an output layer. It was designed for multi-class classification, targeting five distinct classes (DoS, frame fuzzification, RPM, gear spoofing, or normal). We employed a grid search approach to systematically explore and identify the optimal hyperparameters for the ANN classifier. The first layer was the input layer, with an input shape of nine. The next two layers were hidden layers, which are dense layers, each utilizing 16 neurons and the ReLU activation function. The ReLU function was preferred in hidden layers because it introduced non-linearity to the model. The output layer comprised five neurons, corresponding to the number of output labels. For this layer, we employed a softmax activation function, which is suitable for multi-class classification problems. Considering the one-hot encoded labels, we chose the categorical cross-entropy loss function for compiling our model. We optimized the model with the Adam optimizer, maintaining the default values for other parameters. To enhance training efficiency and mitigate overfitting, we integrated an EarlyStopping callback. Table \ref{ANN_Parameters} summarizes the hyper-parameters and their respective values used in tuning the ANN model.

\begin{table}[ht!]
\centering
\caption{ANN Parameters for Multi-classification}
\label{ANN_Parameters}
\begin{tabular}{ll}
\hline
\textbf{Parameter} & \textbf{Value} \\ \hline \\[-.8em]
Epoch & 10    \\ \hline \\[-.8em]
Number of Hidden Layers Neurons & 16    \\ \hline \\[-.8em]
Number of Output Layer neurons & 5  \\ \hline \\[-.8em]
Number of Hidden Layers & 2  \\ \hline \\[-.8em]
Input Layer Activation Function & ReLU \\ \hline \\[-.8em]
Hidden Layer Activation Function & ReLU \\ \hline \\[-.8em]
 Output Layer Activation Function & softmax \\ \hline \\[-.8em]
Optimizer & Adam \\ \hline \\[-.8em]
Batch Size & 256 \\ \hline \\[-.8em]
Shuffle & True \\ \hline \\[-.8em]
Loss Function & categorical\textunderscore crossentropy \\ \hline \\[-.8em]
\end{tabular}
\end{table}

\subsection{LSTM-Autoencoder Hyperparameter Tuning}
In the LSTM-autoencoder model provided, hyperparameters were carefully selected for optimal training. However, before feeding the data into the LSTM-autoencoder, we reshaped the data from a 2D format of (Samples, Features) to a 3D format of (Samples, Time Steps, Features) by using the reshape function. This was because the LSTM requires a 3D input shape. The model operated on sequences of length time\_steps, which was set at 1, meaning each data point was treated as a sequence of its own. Each of these sequences had nine features, which included the CAN ID and eight other data fields from the CAN message. 
This LSTM-autoencoder structure consists of an input layer, two encoder layers, a RepeatVector layer, two decoder layers, and an output layer. The first LSTM layer, which serves as both the input and the first encoder layer, has 128 neurons and is designed to return sequences, enabling the stacking of subsequent LSTM layers. The second LSTM encoder layer, with 64 neurons, does not return sequences. Following the RepeatVector layer, which replicates its input features, the LSTM decoder consists of two LSTM layers with 64 and 128 neurons, respectively. The activation function employed in the input and hidden layers was ReLU, chosen for its rapid and efficient training on large datasets compared to the sigmoid function. To combat overfitting, four dropout layers with a dropout rate of 20\% were instituted after each LSTM layer, randomly deactivating one-fifth of the neurons during every training iteration. We used a batch size of 64 and 100 epochs. Adam optimizer was used along with mean squared error (MSE) loss function, which is known for its application in anomaly detection problems. To enhance training efficiency and to avoid overfitting, we integrated an early-stopping callback and dropout layers.

\begin{table}[ht!]
\centering
\caption{LSTM-Autoencoder Parameters for Binary Classification}
\label{LSTM-Autoencoder_Parameters}
\begin{tabular}{ll}
\hline
\textbf{Parameter} & \textbf{Value} \\ \hline \\[-.8em]
Epoch & 100    \\ \hline \\[-.8em]
Input Layer Activation Function & ReLu  \\ \hline \\[-.8em]
Hidden Layers Activation Function & ReLu  \\ \hline \\[-.8em]
Optimizer & Adam \\ \hline \\[-.8em]
Batch Size & 64 \\ \hline \\[-.8em]
Dropout Rate & 0.2 \\ \hline \\[-.8em]
Shuffle & True \\ \hline \\[-.8em]
Loss Function & MSE\\ \hline \\[-.8em]
\end{tabular}
\end{table}

Table \ref{LSTM-Autoencoder_Parameters} shows the parameters and their respective values in hyper-tuning the LSTM-autoencoder. We trained the LSTM-autoencoder using only the normal data labeled in the dataset as 0, but it was tested using both normal and attack data.

In our anomaly detection method using the LSTM-autoencoder, the model’s capability to reconstruct input data was crucial for detecting anomalies. The reconstruction errors were calculated using the MSE between the reconstructed output and the original data. To delineate a boundary for what qualifies as an anomaly, a threshold was established. Determining the optimal threshold for anomaly detection can be intricate. Through various experiments, we found the most effective threshold is the sum of the average reconstruction error and one standard deviation of these errors from the training set. In simpler terms, the threshold creates a margin above the average error, and any data point with an error exceeding this margin will be considered anomalous. Then, the model can predict anomalies by comparing each test’s reconstruction error to the pre-defined threshold. The threshold is shown in Equation  (\ref{eq:threshold}):

\begin{equation}
\label{eq:threshold}
\text{Threshold} = \mu(\text{train\_errors}) + \sigma(\text{train\_errors})
\end{equation}
In this context, \( \mu \) represents the mean, while \( \sigma \) indicates the standard deviation.

\subsection{Proposed Hierarchical Federated Learning-based Framework}
In this section, we introduce the concept of H-FL, which served as the core technique for enabling our proposed IDS to leverage diverse driving scenarios and behaviors across different locations while ensuring user privacy protection.

Previous researchers have deployed in-vehicle DSs using a traditional centralized learning approach, which involves transmitting large volumes of data to the cloud for training. However, this approach raises privacy concerns and involves high communication overhead and longer response times  \cite{alsamiri2023federated}. Therefore, recent research has shown a growing interest in adopting the FL approach to address these issues.

FL is a unique implementation of a distributed ML approach that involves training a model on edge devices (clients) without transferring the raw data to a central location \cite{mcmahan2017communication}. Thus, FL is an ideal fit for in-vehicle IDSs for several compelling reasons. First, the FL approach preserves data privacy since the learned parameters from local models are sent periodically to the cloud server instead of transmitting the whole row of data. Second, FL allows multiple participants to develop a robust and efficient global model without compromising user data privacy. This feature makes it a better option than non-FL approaches \cite{li2020review}, offering real-time model updates and allowing access to data without contacting the centralized server. Third, FL reduces latency by avoiding transmitting the raw data to a central server. This is crucial, as sending data to a central server can be costly and may impair the effectiveness of IDS deployments \cite{campos2022evaluating}. Fourth, FL’s distributed nature and low complexity make it highly suitable for resource-constrained hardware deployments \cite{driss2022federated}. Based on the guidelines provided by the International Telecommunication Union for IDS in vehicular networks in 2020 \cite{x1375Guidelines}, it is essential that an in-vehicle IDS offers the capability to regularly update its rule sets. Thus, FL makes the IDS adaptive to new, unseen attacks by updating local models with new models trained on new, unseen attacks identified in other clients. This adaptiveness allows the local model to stay up-to-date and more robust against new, unseen attacks. Additionally, since FL involves training models locally on devices and only sharing model updates, it ensures that sensitive data remains private and secure throughout the process. 

However, the FL approach relies solely on one central aggregator, which can introduce performance challenges, particularly due to delays in sharing the model between the central aggregator and numerous devices. Moreover, this aggregator poses a risk as a single point of failure. To address these issues, we decided to deploy the proposed IDS in H-FL, which incorporates a central aggregator with multiple local aggregators instead of having one central aggregator. This approach aimed to overcome the limitations of relying solely on a central aggregator \cite{rana2023hierarchical}.  

\begin{figure*}[h!t]
\centering
    \includegraphics[scale=.11]{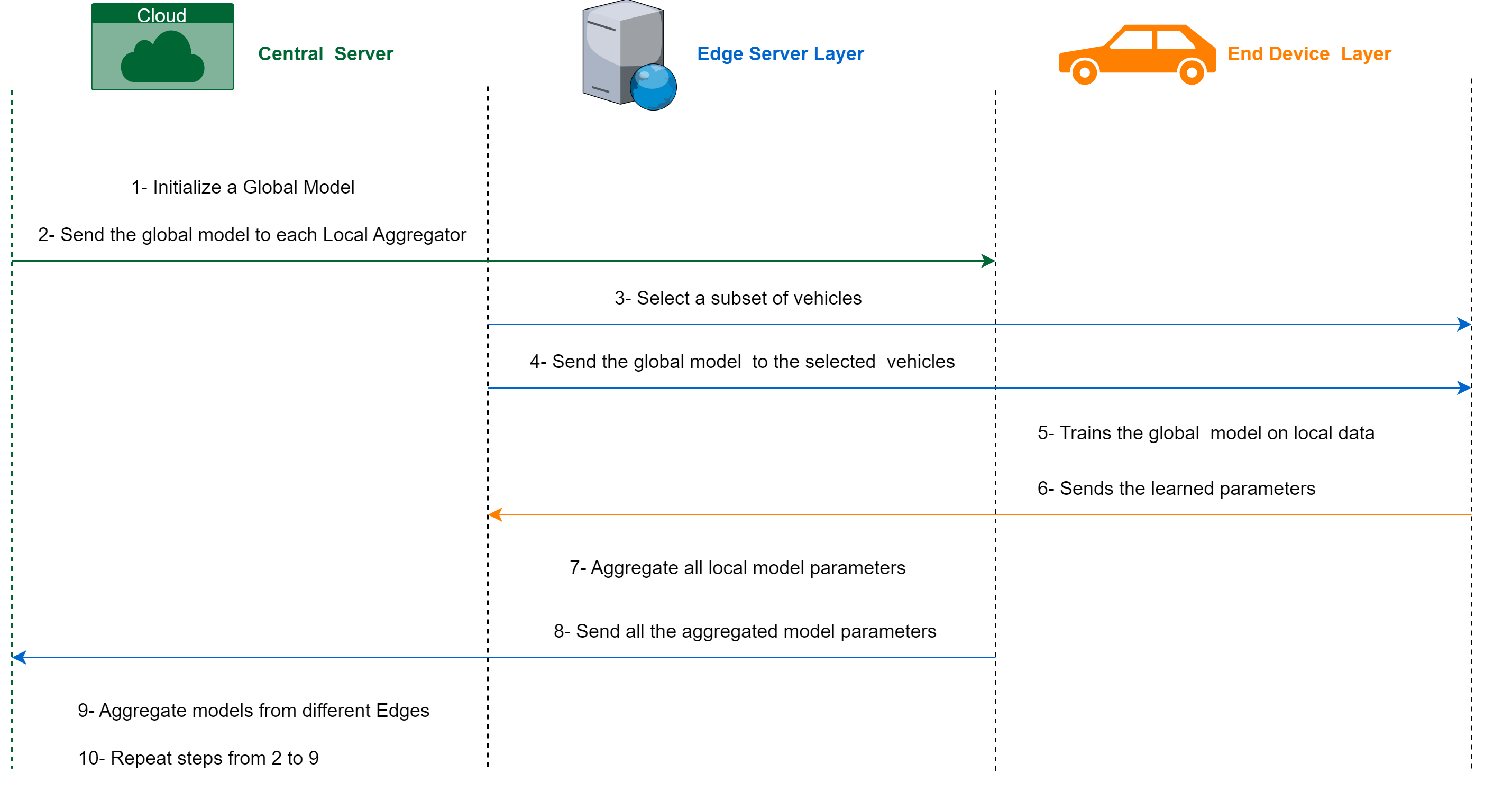}
   \caption{A High-level Overview of H-FL Process}
\label{HFL_Process}
\end{figure*}

The architecture of H-FL consists of three layers: the central server, edge servers, and end devices. The end device layer comprises multiple end devices (vehicles), where each end device has a large amount of local sensor data and a locally installed model. In the edge server layer, multiple local aggregators act as intermediaries for communication between end devices and the central server. Their primary roles include transferring model parameters and aggregating local model parameters. The central server layer, located in the cloud, is accountable for the distribution and continuous updates of the final global model. Figure \ref{HFL_Process} depicts a high-level overview of the H-FL process.

At the beginning of each round, the central server initializes a global model. Then, the central server sends the global model to each local aggregator. Each local aggregator receives the global model, selects a subset of end devices using a selection approach to participate in the training process, and then dispatches the global model to the selected end devices. As CAN bus data can vary significantly between vehicle makes and models, we propose that the local aggregator selects a subset of available end devices (vehicles) based on the similarity in CAN bus data, such as make or model, in specific geographical areas. Grouping vehicles by similarity in CAN bus data ensures that the global model is trained on similar and relevant data, yielding more accurate and reliable results. Once the global model is deployed on each end device, it leverages its local data for training in small batches. After local training, end devices send their trained local model parameters to the corresponding local aggregator. Each aggregator aggregates all the local model parameters/weights using an aggregation algorithm to obtain the edge aggregation model. After all local aggregators have completed the edge model aggregation, they send all the aggregated model parameters to the central server to build a new global model. These aggregated model parameters contribute to refining and improving the global model, which is then sent back to the end devices for continuous training. The central server adjusts the global model parameters based on the aggregated model parameters, preparing for the subsequent round of global training.

The training process in H-FL can be divided into three parts: local training, edge aggregation, and global aggregation. For example, when a local model identifies an anomaly or previously unseen malicious network traffic, it isolates the data for further examination. If this anomaly is determined to be a new, previously unseen attack, a new label is created in the classifier for seen attacks (the first stage in the IDS). In response, the server updates the global model and transmits the updated version to the edge aggregators and then to the selected end device and relevant end devices. After each communication round, the end device returns the model’s parameters to the corresponding edge aggregator and then to the central server, allowing our proposed IDS to adapt. To keep the seen attack detection model effective, seen attack data needs to be updated regularly. It continuously monitors for unseen attacks and updates the supervised classification model upon detection, ensuring its adaptability to new, unseen attacks. This process is repeated until the desired level of performance is obtained.

\section{Results and Evaluations}
\label{Results_Section}

\subsection{Experiment Setup}
The implementation was performed in Google Colab Pro, a web-based editor from Google Research that allows users to write and run arbitrary Python code from the browser. The experimental setup consisted of a 64-bit Ubuntu 20.04.5 LTS operating system, 11th Gen Intel(R) Core(TM) i7-11700 @ 2.50GHz, 31.1GiB RAM, NVIDIA Corporation, and Python 3.9.16 version.

\subsection{Dataset Description}
\label{DatasetDescription}
To assess the performance of our proposed model, we utilized a benchmark dataset published by Song et al. \cite{seo2018gids}. This dataset is extensively used in automotive security research and comprises four types of attacks: DoS, frame fuzzification, engine RPM spoofing, and drive gear spoofing. We selected this dataset because it is based on real-world traffic data rather than simulated data, and it allows us to compare our proposed IDS approach with similar work that uses the same dataset \cite{MTH-IDS2022}. For every CAN message, the dataset provides valuable information, including timestamp, CAN ID, DLC, data field, and flag. The timestamp indicates the precise time when the message was recorded from the startup. The CAN ID plays a crucial role in determining the priority of multiple messages, with lower values being given precedence over higher ones. Moreover, the DLC specifies the data field’s length in bytes, up to 8 bytes. The flag indicates whether the message is normal or an attack. In Table \ref{DatasetFeatures}, we present the dataset’s features, descriptions, and data types.

\begin{table}[ht!]
\centering
\caption{Data Features, Descriptions, and Types}
\label{DatasetFeatures}
\begin{tabular}{lll}
\hline
\textbf{Feature} & \textbf{Describtion} & \textbf{Type} \\ \hline \\[-.8em]
Timestamp & Time & float \\ \hline \\[-.8em]
CAN ID & \begin{tabular}[c]{@{}l@{}} CAN message identifier \end{tabular} & hexadecimal \\ \hline \\[-.8em]
DLC & \begin{tabular}[c]{@{}l@{}}The size of the data field, measured\\ in bytes\end{tabular} & integer \\ \hline \\[-.8em]
Data & Payload (64-bit) & hexadecimal \\ \hline \\[-.8em]
Flag &\begin{tabular}[c]{@{}l@{}}T or R, \\ T: Attack, R: Normal\end{tabular}  & string \\ \hline \\[-.8em]
\end{tabular}
\end{table}

To select the most suitable algorithms, we examined the CAN bus data from Car Hacking Dataset \cite{seo2018gids}, ensuring a comprehensive understanding of both normal and abnormal behaviors. Attacks include:

\textbf{DoS Attack}: Inject many ZERO values into every CAN bus ID and payload at 0.3 millisecond intervals. This dominates the BUS, causing legitimate messages to be delayed or blocked.

\textbf{Frame Fuzzification Attack}: Messages are randomly injected into the CAN bus. There are two types of frame fuzzification attacks present in this dataset:
\begin{enumerate}
    \item Injecting random IDs (not seen before).
    \item Injecting IDs that appear legitimate but have a different payload.
\end{enumerate}
During a frame fuzzification attack, an adversary might expect some valid CAN messages to inadvertently cause a malfunction in the target vehicle. It is presumed that the adversary has no prior knowledge of the in-vehicle communication of the target vehicle. Thus, the adversary injects messages with random CAN IDs and payloads. This implies that both the CAN IDs observed in normal traffic and those not seen before can be included in a frame fuzzification attack.

\textbf{RPM Spoofing Attack}: This spoofing attack specifically targets RPM CAN ID: 790, aiming to inject fabricated messages to control various functions. The legitimate data payload is different from the fabricated messages.

\textbf{Gear Spoofing Attack}: This spoofing technique targets the Gear CAN ID: 1087, attempting to inject fabricated messages to control functions. While the fabricated messages resemble normal ones, they are not identical. For illustration purposes, Table \ref{DatasetData} presents examples of both normal data and various kinds of attack data. Data represented in black indicates normal data, while data in red signifies attack data.

\begin{table}[]
\centering
\caption{Normal and Attack data in Car Hacking Dataset}
\label{DatasetData}
\setlength{\tabcolsep}{5pt} 
\begin{tabular}{cccccccccc}
\hline
\textbf{CAN ID} & \textbf{D0} & \textbf{D1} & \textbf{D2} & \textbf{D3} & \textbf{D4} & \textbf{D5} & \textbf{D6} & \textbf{D7} & \textbf{Class} \\  \hline
880 & 0 & 64 & 96 & 255 & 120 & 0  & 8  & 0  & Normal \\ \hline
{\color[HTML]{FE0000} 0}  & {\color[HTML]{FE0000} 0} & {\color[HTML]{FE0000} 0}  & {\color[HTML]{FE0000} 0} & {\color[HTML]{FE0000} 0}   & {\color[HTML]{FE0000} 0}  & {\color[HTML]{FE0000} 0}   & {\color[HTML]{FE0000} 0}  & {\color[HTML]{FE0000} 0}  & DoS    \\ \hline
{\color[HTML]{FE0000} 55} & {\color[HTML]{FE0000} 0} & {\color[HTML]{FE0000} 1}  & {\color[HTML]{FE0000} 1} & {\color[HTML]{FE0000} 0}   & {\color[HTML]{FE0000} 11} & {\color[HTML]{FE0000} 22}  & {\color[HTML]{FE0000} 0}  & {\color[HTML]{FE0000} 1}  & Frame fuzzification  \\ \hline
880                       & {\color[HTML]{FE0000} 0} & {\color[HTML]{FE0000} 0}  & {\color[HTML]{FE0000} 0} & {\color[HTML]{FE0000} 255} & {\color[HTML]{FE0000} 0}  & {\color[HTML]{FE0000} 0}   & {\color[HTML]{FE0000} 10} & {\color[HTML]{FE0000} 0}  & Frame fuzzification  \\ \hline
790                       & 0                        & 30                        & 40                       & 0                          & 0                         & 13                         & 0                         & 9                         & Normal \\ \hline
790                       & {\color[HTML]{FE0000} 0} & {\color[HTML]{FE0000} 8}  & {\color[HTML]{FE0000} 8} & {\color[HTML]{FE0000} 0}   & {\color[HTML]{FE0000} 0}  & {\color[HTML]{FE0000} 0}   & {\color[HTML]{FE0000} 0}  & {\color[HTML]{FE0000} 11} & RPM    \\ \hline
1087                      & 1                        & 2                         & 1                        & 0                          & 0                         & 1                          & 240                       & 7                         & Normal \\ \hline
1087                      & {\color[HTML]{FE0000} 0} & {\color[HTML]{FE0000} 22} & {\color[HTML]{FE0000} 1} & {\color[HTML]{FE0000} 180} & {\color[HTML]{FE0000} 0}  & {\color[HTML]{FE0000} 130} & {\color[HTML]{FE0000} 0}  & {\color[HTML]{FE0000} 33} & Gear   \\ \hline
\end{tabular}
\end{table}

\subsection{Evaluation Metrics and Performance Evaluation}
To assess the robustness of the proposed IDS, we considered various performance metrics such as accuracy (Acc), F1-score (F1), precision (Pre), recall (Rec)—or, as it is called, DR—and FAR. Metrics were determined based on true positive (TP), true negative (TN), FP, and FN values. We used the following equations to calculate the metrics used:

\begin{equation}
Acc=\frac{TP+TN}{TP+TN+FP + FN}
\end{equation}

\begin{equation}
F1 =2 \times \frac{Pre \times Rec}{Pre+Rec}
\end{equation}

\begin{equation}
Pre=\frac{TP}{TP + FP}
\end{equation}

\begin{equation}
Rec=\frac{TP}{TP + FN}
\end{equation}

\begin{equation}
FAR=\frac{FP}{TN + FP}
\end{equation}

\subsection{Performance Results and Analysis of Seen and Unseen Attacks Detection}
This subsection summarizes the results of our proposed IDS in detecting seen and unseen attacks and provides an analysis of these findings.

\begin{figure}[ht!]
\centering
    \includegraphics[scale=.65]{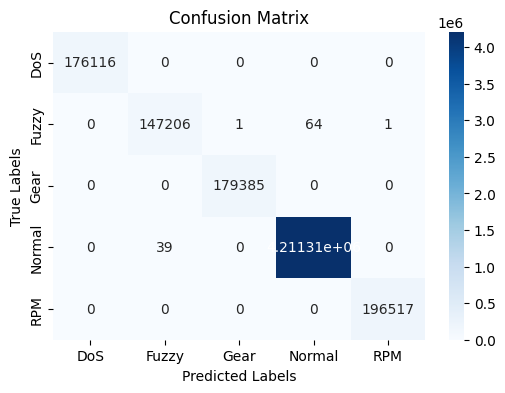}
    \caption{ANN Multiclass Confusion Matrix}
    \label{ANNConfusionMatrix1}
\end{figure}

\begin{figure}[ht!]
\centering
    \includegraphics[scale=.55]{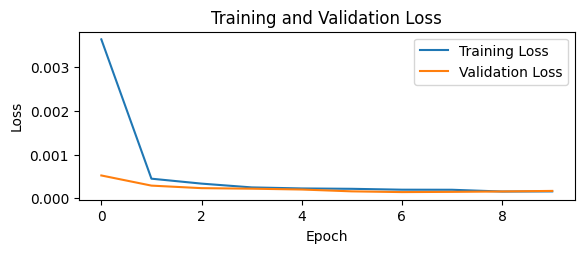}
    \caption{ANN Training and validation performance}
    \label{ANNLoss}
\end{figure}

Starting with the results for seen attack detection in the first model, the ANN was trained and tested on a labeled dataset that included normal data and four types of attacks: DoS, frame fuzzification, RPM, and gear spoofing. The performance of the ANN model is detailed in Table \ref{ANN_results}, which shows that the ANN model consistently achieved impressive accuracy and F1-scores exceeding 99\% in accurately classifying various types of seen attacks. Additionally, the table highlights the model’s precision, its DR, and FAR for normal data and each attack category, demonstrating the model’s high reliability and effectiveness in distinguishing between normal and various types of attacks. Figure \ref{ANNConfusionMatrix1} displays the multiclass confusion matrix (CM) of the ANN model, which illustrates the model’s capability to classify test data into multiple attack categories. Moreover, as depicted in Figure \ref{ANNLoss}, the training and validation losses decreased and converged over time, which indicates that the model was learning effectively and generalizing well without overfitting.

\begin{table}[ht!]
\caption{Performance Evaluation of ANN on Seen Attacks Detection}
\centering 
\label{ANN_results} 
\begin{tabular}{clllll} 
\hline 
\multicolumn{1}{l}{\textbf{Attack}} & \textbf{Acc (\%)} & \textbf{F1} & \textbf{Pre} & \textbf{DR(\%)} & \textbf{FAR(\%)} \\ \hline 
DoS & 99.99 & 1.00 & 1.00 & 100 & 0.0\\ \hline 
Frame Fuzzification  & 99.99 & 0.99 & 0.99 & 99.95 & 0.0005 \\ \hline 
Gear &99.99  & 1.00 & 0.100 & 100 & 0.0 \\ \hline
RPM & 99.99 & 0.99 & 0.99 & 100 & 0.0 \\ \hline 
Normal & 99.99  &0.99  & 0.99  & 99.99 & 0.012 \\ \hline 
\end{tabular} 
\end{table}

\begin{figure}[ht!]
    \centering
    \begin{subfigure}[b]{0.45\textwidth}
        \centering
        \includegraphics[scale=0.55]{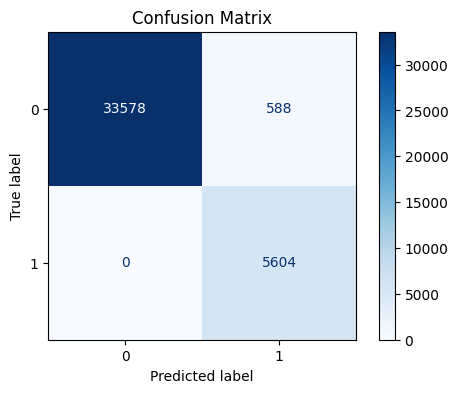}
        \caption{SMOT Sampeling}
        \label{LSTMConfusionMatrix_SMOT}
    \end{subfigure}
    \hfill
    \begin{subfigure}[b]{0.45\textwidth}
        \centering
        \includegraphics[scale=0.55]{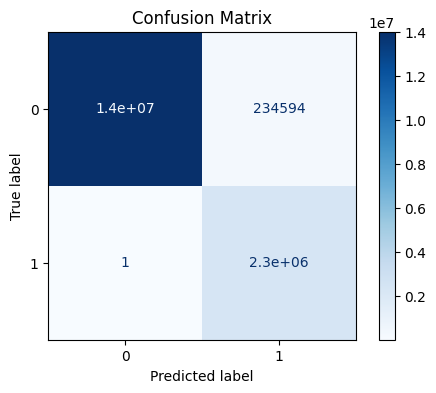} 
        \caption{All Remaining Testing Data}
        \label{LSTMConfusionMatrix_ALL}
    \end{subfigure}
    \caption{LSTM-autoencoder Binary Confusion Matrices}
    \label{LSTMConfusionMatrix}
\end{figure}

To evaluate the model’s ability to detect new, unseen attacks, we trained the LSTM-autoencoder on a sample of normal data and then tested it on the remaining dataset. As shown in Table \ref{Unseen_Attack_Results}, the LSTM-autoencoder performed well, with an overall accuracy of 98.59\%, an F1-score of 0.95, a DR of 99.99\%, and a precision of 0.91 across all types of unseen attacks. Despite generating approximately 0.016\% FAR, these metrics underscore the model’s efficacy in detecting new, unseen attacks. For each type of unseen attack, the model successfully detected all attacks with a DR between 99.99\% and 100\%, accuracy exceeding 98\%, and low FAR of 0.016\%. However, the F1-score varied from 0.81 to 0.85 for specific types of unseen attacks. Figure \ref{LSTMConfusionMatrix} shows the binary CM of the LSTM-autoencoder model, illustrating its performance in classifying the test data into normal (0) and anomaly (1) classes. Figure \ref{LSTMConfusionMatrix_SMOT} shows the CM on the test samples after applying SMOT sampling, as described in Section \ref{Data_Preprocessing}, while Figure \ref{LSTMConfusionMatrix_ALL} displays the CM for the remaining test set in the dataset. From the results, it is clear that the model successfully detected all unseen attacks, even when it had not been trained on them before.

\begin{table}[ht!]
\centering
\caption{Detection Results for Unseen Attacks}
\label{Unseen_Attack_Results}
\begin{tabular}{cccccc}
\hline
\textbf{Unseen Attack} & 
\textbf{F1} & \textbf{Pre} & \textbf{Acc (\%)} & \textbf{DR (\%)} & \textbf{FAR (\%)}\\ \hline
DoS & 0.83  & 0.71   & 98.42   & 100 & 0.016 \\ 
Frame Fuzzification & 0.81 & 0.68  & 98.41  & 99.99  & 0.016 \\ 
Gear & 0.84   & 0.72  & 98.42   &  100  &  0.016  \\ 
RPM & 0.85  &  0.74 & 98.42 & 100  & 0.016  \\
\rowcolor{yellow!50}
\textbf{All} & 0.95  & 0.91  & 98.59  & 99.99 & 0.016  \\ \hline
\end{tabular}
\end{table}

Our results indicate that the ANN successfully detected and classified attacks by type. This capability is crucial, as identifying the specific attack type aids in selecting appropriate countermeasures and conducting post-attack analysis \cite{zhao2022can}. As shown in  Table \ref{Unseen_Attack_Results}, the model showed different F1-scores for unseen attack detection when testing each attack individually compared to testing all attacks together. The F1-score varied from approximately 0.81 to 0.95. This is because the number of TPs significantly increased when combined, while the number of FPs remained similar. Consequently, this increase in TPs improved both precision and recall, resulting in a higher F1-score. 
Moreover, from the high DR and the constant FAR across all results, we can observe that while the model was able to detect all unseen attacks, it mistakenly classified some normal data as attacks, accounting for 234,994 FPs. One reason for this could be that the model was trained on a small sample of normal data. Although having 234,994 FPs out of approximately 14,000,000 is a good result, it should be further improved upon in such a critical application. Nevertheless, our findings reveal promising results in detecting unseen attacks.

\subsection{Model Complexity}
This section discusses model complexity in terms of model size (in megabytes, MB) and the number of trainable parameters. When designing in-vehicle IDS solutions, it is essential to consider the deployment requirements \cite{lokman2019intrusion}. The development and deployment of IDSs are significantly impacted by the constraints of ECU in-vehicle networks, which include limited memory storage, computing power, and bandwidth \cite{rajapaksha2023ai}. In pursuit of optimal results, we have simplified the model architecture to minimize its size. The proposed IDS achieves this reduction by employing a straightforward architecture with a minimal number of layers and neurons in both models, as well as the dropout regularization technique in the LSTM-autoencoder. Through careful experimentation with hyperparameters, we optimized the model’s efficiency, resulting in a lightweight architecture. The sizes of the ANN and LSTM-autoencoder models were calculated to be 0.030 MB and 2.95 MB, respectively. Moreover, the number of parameters significantly influences the model’s training and testing time. In theory, a model with fewer parameters will train and test more quickly \cite{hoang2022detecting}. The ANN model has 517 trainable parameters, while the LSTM-autoencoder has 253,065, making a combined total of 253,582 trainable parameters.

\subsection{Comparison with Existing Studies}
This model is compared with recent work in \cite{MTH-IDS2022}, since they used the same dataset and a similar approach and features. Regarding the seen attack detection results, both have a high DR with an F1-score of 0.99. 
However, in detecting unseen attacks, even though it is difficult to obtain a fair comparison, we made an effort to make the best possible comparison. To do so, we used the same numbers of testing instances for attack and normal instances as were used in \cite{MTH-IDS2022}. Results in Table \ref{Comparison} show that our model outperformed the results in \cite{MTH-IDS2022}, with a higher DR and lower FAR. For the F1-score, our average was 0.95, while their model achieved a slightly higher score of 0.96. However, the F1-score for unseen attacks in \cite{MTH-IDS2022} was initially around 0.83, and they improved the result to 0.96 by implementing two biased classifiers after the unsupervised model, achieving a DR of around 93\%. Training these biased classifiers on FPs and FNs, however, transforms the model from being purely unsupervised. 
Although \cite{hoang2022detecting} and \cite{seo2018gids} used the same dataset as ours, we did not compare our detection results with theirs because they relied solely on the CAN ID feature to build their models.

Most previous papers do not state the model sizes, except \cite{MTH-IDS2022} and \cite{hoang2022detecting}, so we compared our model with theirs. As depicted in Table \ref{Model_Size_Comparison} the model size in \cite{MTH-IDS2022} for the two models is 2.61 MB, and the total size of our models is 2.98 MB, showing that our model is nearly in the same range even though we used DL, which is considered more resource-intensive than traditional ML. These sizes are notably below the typical memory capacity of vehicle-level machines, which can exceed 1 GB of RAM \cite{MTH-IDS2022}. Moreover, Table \ref{Model_Size_Comparison} shows that our trainable parameters represent approximately an 88.2\% reduction compared to the number of trainable parameters in \cite{hoang2022detecting}, even though they only used one feature, which is the CAN ID.

Therefore, the experimental results confirm that our proposed DL-based, in-vehicle IDS is highly efficient and can effectively detect various types of seen and unseen cyberattacks. Additionally, its lightweight design makes it feasible for real-world deployment.

\begin{table}[]
\centering
\caption{Comparison with Existing Work}
\label{Comparison}
\begin{tabular}{cllllll}
\hline
\multicolumn{1}{l}{\multirow{2}{*}{\textbf{Unseen Attack}}} & \multicolumn{3}{c}{\textbf{Ours}} & \multicolumn{3}{c}{\textbf{MTHIDS \cite{MTH-IDS2022}}} \\ \cline{2-7} 
\multicolumn{1}{l}{}                                         & \textbf{DR (\%)}           & \textbf{FAR (\%)}       & \textbf{F1 }     & \textbf{DR (\%) }     & \textbf{FAR (\%) }     & \textbf{F1}      \\ \hline
DoS &   100   &    0.016   &    0.95  &    100  &  0.0    &   1.0    \\ \hline
Frame Fuzzification  &  100  &   0.016    &  0.94    &   73   &   0.057    & 0.84        \\ \hline
Gear   &   100   &   0.016     &   0.95   &     100    &   0.45    &   0.99   \\ \hline
RPM   &   100  &   0.016  &   0.95  &   100  &  0.003  &   0.99  \\ \hline
\rowcolor{yellow!50}
\textbf{Average}  &    100  &    0.016     &     0.95    &  93.7   &   0.128    &   0.96  \\ \hline

\end{tabular}
\end{table}

\begin{table}[]
\centering
\caption{Model Size comparison}
\label{Model_Size_Comparison}
\begin{tabular}{ccc}
\hline
\textbf{Model}  & \multicolumn{1}{l}{\textbf{MB}} & \multicolumn{1}{l}{\textbf{Trainable parameters}} \\ \hline
MTHIDS \cite{MTH-IDS2022}   & 2.61     & -     \\ \hline
AE- GAN \cite{hoang2022detecting} & -  & 2.15 million    \\ \hline
\textbf{Ours}    & 2.98  & 253,582      \\ \hline
\end{tabular}
\end{table}

\section{Discussion}
\label{Discussion}
\begin{figure*}[ht!]
\centering
    \includegraphics[scale=.12]{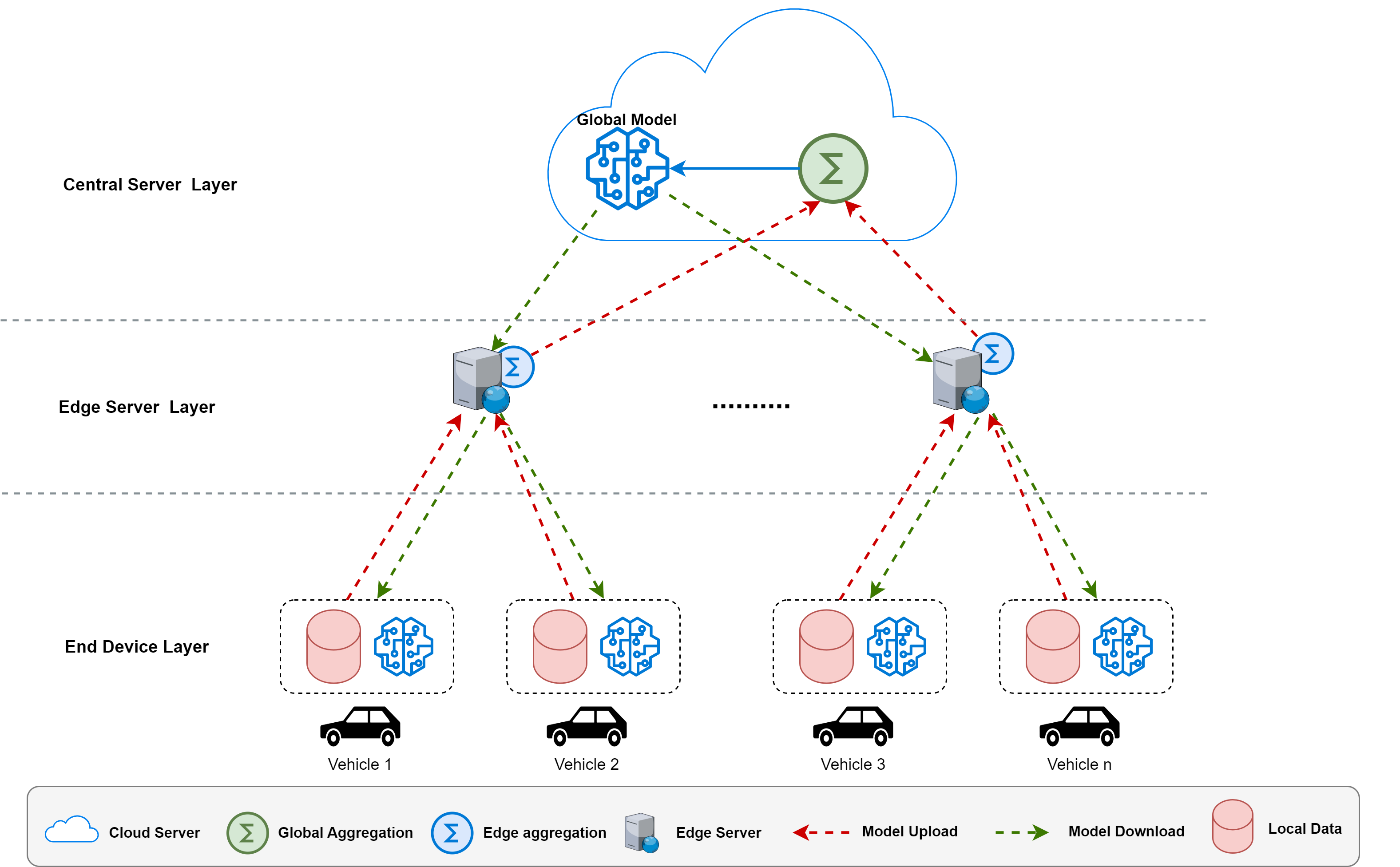}
    \caption{Framework of the proposed Hierarchical Federated Learning Method}
    \label{fig:HFL-Framework}
\end{figure*}

Our analysis revealed several key findings that contribute to the understanding and development of in-vehicle IDSs:
\begin{itemize}
    \item Hybrid IDSs, such as our proposed IDS, can be a robust solution that not only addresses current threats but also prepares for future ones. Moreover, the order of each approach is important. For example, we adopted the seen attack detection before anomaly detection approach for two reasons: to quickly detect any seen attacks and to double-check the normal data in case an attack bypasses the first model.

   \item When designing an in-vehicle IDS, several critical decisions should be made during the design phase. One such decision is to include both CAN ID and payload data without feature selection, as attackers might exploit any neglected features in the future\cite{li2014feature,zhang2015adversarial}.

     \item Our analysis of CAN bus data shows that each CAN ID has unique data patterns, so important features for one ID may not be relevant for another. This variability makes it impractical to select a consistent set of important features for all CAN IDs, indicating the need to customize feature selection to improve model performance and generalization.
     
   \item The most important finding is that our results prove that DL algorithms can improve the performance of an IDS while meeting the model size requirements in resource-constrained environments.

   \item Theoretically, and based on the literature review, an H-FL architecture, which adds an edge layer between the central server and the vehicles, can overcome several challenges in the traditional FL architecture that consists only of a server and vehicles. Figure \ref{fig:HFL-Framework} depicts the theoretical framework of the proposed H-FL.
\end{itemize}

\section{Conclusions and Future Directions}
\label{Conclusion}
The aim of this paper was to propose a robust and lightweight multi-stage IDS designed for in-vehicle network security that is capable of detecting both seen and novel attacks. Our IDS addresses the limitations of existing solutions by utilizing a hybrid approach and advanced DL algorithms. To further enhance our IDS and leverage diverse driving behaviors while preserving data privacy, we have proposed a theoretical framework for deploying our IDS in an H-FL environment. We evaluated the performance of our IDS using a real-world dataset containing various cyberattacks, including DoS, frame fuzzification, RPM, and gear spoofing. Experimental results demonstrate that the ANN model effectively classifies seen attacks with an outstanding F1-score of 0.99. Simultaneously, the LSTM-autoencoder model excels at detecting novel attacks, achieving an F1-score of over 0.95 and a DR of 99.99\% with minimal false alarms. Overall, our proposed IDS effectively detects both seen and novel attacks within in-vehicle networks and continually updates its knowledge by identifying new, previously unseen attacks, ensuring ongoing improvement over time. Additionally, our IDS is designed to be lightweight, making it suitable for real-world deployment. By detecting both seen and novel attacks, our IDS not only addresses current threats but also prepares for future ones. For future work, we plan to deploy our proposed IDS in a realistic H-FL environment and evaluate its performance.

Although our proposed IDS shows promising results in detecting both seen and novel attacks while maintaining a compact model size, it has certain limitations. Our IDS has been trained and evaluated within limited driving scenarios, requiring extensive datasets to model normal behavior accurately. This limitation suggests potential areas for enhancement, which can be addressed through the following future directions:
\begin{itemize}
    \item Future research could explore streaming learning, allowing the model to dynamically adjust in real time within the vehicle to adapt to various driving conditions, thus enhancing detection accuracy. 
  
   \item FL can effectively combine models derived from different driving scenarios and vehicle states, greatly improving in-vehicle IDS performance while protecting data privacy and reducing latency \cite{zhang2023federated}. Thus, exploring the field of FL and addressing its challenges, such as data heterogeneity \cite{li2020federated} and secure communication \cite{barati2016rdtp}, can be identified as future trends in in-vehicle IDS research.

   \item Another crucial future direction is protecting in-vehicle IDSs from adversarial attacks, as a recent study \cite{aloraini2024adversarial} has highlighted their vulnerability. Protecting in-vehicle IDSs from adversarial attacks and adapting solutions from other domains could provide valuable insights and improvements to current in-vehicle IDSs.
\end{itemize}

\bibliographystyle{unsrtnat}
\bibliography{references}

\end{document}